\documentclass[aps,prb,twocolumn]{revtex4} 

\usepackage{graphicx}
\usepackage{dcolumn}
\usepackage{bm}
\usepackage{amsmath}  

\newcommand{\comment}[1]{}

\begin{document}

\title{Bose-Einstein Condensation and the Lambda Transition
for Interacting Lennard-Jones Helium-4}


\author{Phil Attard}
\affiliation{ {\tt phil.attard1@gmail.com}  \today}
\noindent {\tt  Projects/QSM25/lambda/lambda.tex}


\begin{abstract}
An introduction to Bose-Einstein condensation
and the $\lambda$-transition is given.
Results of quantum loop Monte Carlo simulations are presented
for interacting Lennard-Jones helium-4.
The optimum condensation fraction
is found by minimizing the constrained free energy.
The results show that approaching the transition
the growth of pure position permutation loops
and the consequent divergence of the heat capacity
are enabled by the suppression of condensation
and consequently of superfluidity.
Condensation and superfluidity emerge
at the peak of the heat capacity
due to mixed position permutation chains.
\end{abstract}


\maketitle

%
\section{Introduction}
\setcounter{equation}{0} \setcounter{subsubsection}{0}
\renewcommand{\theequation}{\arabic{section}.\arabic{equation}}
%

In the fields of superfluidity and the $\lambda$-transition
there are several important questions that have received little attention.
First, what suppresses Bose-Einstein condensation
above the $\lambda$-transition?
After all, if, as conventional wisdom asserts,
Bose-Einstein condensation is bosons in the ground energy state,
its very hard to see why there are \emph{no}
bosons in the ground energy state above the $\lambda$-transition.
Yet this must be the case if superfluidity is carried by condensed bosons
because superfluidity is not observed above the $\lambda$-transition.

Second, why does the heat capacity diverge,
and why does it do so over such a narrow temperature range
(about 0.4\,K)?
In the widely accepted non-interacting boson analysis
of Bose-Einstein condensation and the $\lambda$-transition,
the peak of the heat capacity is finite
(Attard 2025a, Le Bellac \emph{et al.}\ 2004,
F. London 1938, Pathria 1972).
Thus the measured divergence must be due to interactions
between the atoms.
It remains to give the molecular structure causing it
and to explain why it occurs over such a narrow range.

And third, why does this divergence in the heat capacity coincide
with the onset of superfluidity?
It is as if the divergence itself catalyzes condensation
and overcomes the previous suppression,
but it is difficult to understand this at the molecular level.

This paper answers these questions
using quantum loop Monte Carlo simulations of Lennard-Jones $^4$He,
guided by existing experimental observations.

\subsection{Background}

I take it as axiomatic
\begin{itemize}
\item
that the $\lambda$-transition
and superfluidity in liquid helium-4
are due to Bose-Einstein condensation

\item
that a condensed boson
is one in a highly occupied momentum state.
\end{itemize}

The $\lambda$-transition is signified by a spike
in the heat capacity of saturated liquid $^4$He at 2.2\,K.
The experimental evidence
is that on the liquid saturation curve
the energy, the density, and the shear viscosity
are continuous functions of temperature at the $\lambda$-transition;
the density and the shear viscosity
have a discontinuity in their first temperature derivative
(Donnelly and Barenghi 1998).
The heat capacity has an integrable  divergence at the $\lambda$-transition
(Lipa \emph{et al.}\ 1996).
Superfluid flow occurs in thin films and capillaries
immediately below the $\lambda$-transition, but not above it.

In so far as superfluidity is due to Bose-Einstein condensation,
the first axiom,
this last fact indicates that there is no condensation
above the $\lambda$-transition,
and that there is condensation below the $\lambda$-transition
sufficient for the observed flows.
It is conventionally understood
that the condensation transition is continuous,
with condensation increasing from zero at the transition itself.
This is evidenced by the continuity in energy, density, and shear viscosity,
and it is also predicted by the non-interacting boson model
(F. London 1938, Pathria 1972).
However it remains to reconcile this conclusion
with the discontinuous appearance of superfluidity
(flow in thin films and capillaries,
and the absence of boiling in the undersaturated liquid)
at the transition.
In any case, that the condensation is macroscopic is confirmed
by the differences on either side of the transition
in the behavior of the energy and heat capacity,
and also the number (in a given volume),
and the flow,
since these are all extensive thermodynamic variables.

It has always been assumed,
ever since Einstein (1924, 1925) first asserted it,
that Bose-Einstein condensation is solely into the ground energy state.
However, the experimental fact
that there is no latent heat at the $\lambda$-transition
argues against this since the appearance of a macroscopic number
of bosons with zero energy would create a sudden macroscopic
change in energy.
Rather, the absence of latent heat suggests
that the number of bosons in any given energy range
is the same before and after condensation.
This is consistent with a transition
from singly to multiply occupied momentum states,
as in the second axiom above,
since this can be accomplished in any kinetic energy neighborhood
by increasing the momenta of some bosons while decreasing that of others
without changing the total energy.

A caveat to this deduction is that
if the amount of macroscopic condensation
is zero at the $\lambda$-transition,
then it might be possible for the condensed bosons
to occupy solely the ground energy state
without a macroscopic energy discontinuity.
However
the full weight of evidence is that condensation occurs
in multiple low-lying momentum states.
For example,
for non-interacting (ideal) bosons the occupancy of the ground energy state
is an intensive thermodynamic variable (Attard 2025)
(this conclusion also holds on general thermodynamic grounds
for interacting particles),
whereas the experimental evidence is that the $\lambda$-transition
is a transition in an extensive thermodynamic variable.
Also, bosons involved in superfluid flow necessarily have non-zero velocity,
which by definition means that they cannot be in the ground energy state.

The $\lambda$-transition and superfluidity
have been explored at the molecular level
within the framework of  quantum statistical mechanics
(Attard 2018, 2021, 2025a).
Detailed mathematical analysis and computer simulations
have revealed the role played by interactions between the $^4$He atoms,
which were missing in earlier modeling with ideal bosons.

There is an important difference between classical statistical mechanics
(Attard 2002)
and quantum statistical mechanics formulated in classical phase space
(Attard  2021)
that is directly relevant to Bose-Einstein condensation
and the $\lambda$-transition.
In classical statistical mechanics,
the entropy of a macrostate is the logarithm of the weighted sum
of the molecular configurations giving that macrostate.
This is, in essence, Boltzmann's original explanation of entropy.
Hence the partition function, which is essentially the total entropy,
is an integral over phase space.
In quantum mechanics,
 the wave function for bosons must be symmetrized
by summing over all permutations.
Each permutation of a configuration is a new configuration.
In quantum statistical mechanics,
in addition to the integral over phase space,
one also has to sum over all permutations to obtain the total entropy.

In phase space each boson has a position and a momentum,
with the latter most usefully quantized
in the vicinity of and below the $\lambda$-transition.
A permutation swaps the momenta of different bosons.
Permuting the momenta of condensed bosons all in the same momentum state
leaves the configuration unchanged.
Such permutations have unit weight,
and the occupation entropy,
which is the logarithm of the number of these permutations,
is what ultimately drives Bose-Einstein condensation.

Using quantum statistical mechanics
to explain the $\lambda$-transition
means identifying both the regions of phase space
and the particular types of permutations that dominate.
Most permutations give rise to highly oscillatory weights
that cancel with small changes in the position or momentum configuration.
The dominant permutations give real, positive weights
that are insensitive to small changes in configuration.
The specific types of permutation that significantly contribute
to the total entropy depend upon the thermodynamic state of the system.

At higher temperatures
the number of accessible momentum states far exceeds the number of bosons.
In this case every boson is the sole occupant of its own momentum state,
and there is no condensation.
All permutations give highly oscillatory
and therefore canceling weight factors,
and so only the identity permutation gives
a non-zero phase space weight.
This is the classical regime.

The number of accessible momentum states decreases
with decreasing temperature.
Obviously there comes a temperature when for a fixed number of bosons
it is impossible to prevent
the multiple occupancy of momentum states.
This is the condensed or quantum regime.
Approaching the transition to the quantum regime,
and deep inside it, there are classes of permutations
that give real, positive weights,
and these therefore must be taken into account.

On the high temperature side of the $\lambda$-transition
the divergence of the heat capacity
is due to the growth in number and size of position permutation loops.
A position loop is a cyclic permutation
around a ring of bosons with successive neighbors in close spatial proximity.
These loops begin to form when
the Gaussian with thermal wavelength width overlaps with
the first peak in the pair distribution function:
the thermal wavelength increases with decreasing temperature,
and the first peak is located at about the diameter of the $^4$He atom,
and increases in height with decreasing temperature.
These position permutation loops
are a subset of all the possible permutations:
their real positive weight,
which is less than unity for each loop,
results from averaging over momenta
and it is insensitive to small changes in position configuration.
The  loops arise from wave function symmetrization,
which itself sums over all possible permutations of the bosons.
The sum of the weights attached to these
position permutation loops contributes to the entropy of the system.

As mentioned,
permutations amongst condensed bosons all in the same momentum state
have unit weight and begin to dominate
below the $\lambda$-transition.
The total number of permutations due to such bosons is just
the product of the factorials of the momentum state occupancy,
and this also contributes to the entropy of the system.
These particular permutations can be constructed from what I call
momentum permutation loops,
in contrast to position permutation loops.
The sum of all momentum loops constructed from bosons in a momentum state
is the factorial of the occupancy of that state.
In practical terms calculations are performed with occupancies
rather than with momentum loops.
But in conceptual terms the fact that loops must be disjoint
underscores the point that an individual boson cannot belong
to both a momentum and a position permutation loop
in a single permutation.
Whereas position permutation loops are compact
and localized at the molecular level in position space,
momentum loops are non-localized:
bosons in the same momentum state can be permuted with unit weight
even when they are separated by macroscopic distances.

Position and momentum permutation loops are the two most important
ways to formulate wave function symmetrization,
and they appear sufficient to account for most aspects
of the $\lambda$-transition and superfluidity.
It makes sense therefore in the first instance
to factorize the phase space weight
that accounts for wave function symmetrization
into pure position and momentum loop symmetrization functions.
The loops that form any permutation must be disjoint,
which is to say that a particular boson can be in only one loop
in any given permutation.
Therefore,
in any configuration only currently condensed bosons
contribute to the momentum loop symmetrization function,
and only currently uncondensed bosons
(and condensed bosons not currently involved in momentum permutation loops)
contribute to the position loop symmetrization function.
Hence the two types of loops compete for the available bosons
and shift the equilibrium between them.
Which one dominates depends upon the associated entropy.

This is the simplest binary description:
bosons are either condensed or uncondensed,
and loops are either pure position or pure momentum.
In a more nuanced picture,
which is necessary to explain the full range of observed phenomena,
it is only a subset of the condensed bosons,
namely those in the most highly occupied momentum states,
that are entropically favored by permutations within their state
at the expense of the entropy lost
by destroying the position permutation loops
from which they are excluded.
Also mixed chains, which are open position loops
with a condensed boson at the head and uncondensed bosons forming the tail,
appear necessary to account fully
for the far side of the $\lambda$-transition.
These are discussed in more detail below and in the text.

The number and size of position permutation loops
with significant weight grow
approaching the $\lambda$-transition from the high temperature side
due to the increase of the thermal wavelength
and of the peak of the pair distribution function.
At the molecular level position loops are structurally organized
in position space.
In the binary picture,
the transformation of an uncondensed to a condensed boson
with momentum loop permutations
destroys all the position loops it was part of,
and the system loses the associated position loop symmetrization entropy.
On the high temperature approach to the $\lambda$-transition
the available position symmetrization entropy per boson is higher
than the available occupation entropy for condensed bosons,
which means that position loops suppress condensation in this regime.
The computational results below confirm
this to be true immediately preceding the $\lambda$-transition
in the regime where the heat capacity is diverging.
This is consistent with the experimental evidence
for the absence of superfluidity above the $\lambda$-transition,
namely that it implies the absence of condensed bosons.

This is an important difference between the ideal
and the interacting boson models of the $\lambda$-transition.
The ideal boson model has no positional structure or position loops
and no physical mechanism to suppress condensation.
It is simply asserted in the ideal boson model
that the ground state is unoccupied above the transition,
and that condensation occurs, and only occurs,
when the excited states become full within the ideal boson model
(F. London 1938, Pathria 1972).
For non-interacting bosons the peak in the heat capacity is finite,
$C_V^\mathrm{id}/Nk_\mathrm{B} = 1.925$
(Attard 2025a, F. London 1938, Pathria 1972).
For real $^4$He (Lipa \emph{et al.}\ 1996)
and for Lennard-Jones $^4$He (Attard 2025a)
the peak in the heat capacity is infinite.
In the calculations for interacting bosons
the divergence of the heat capacity
is due to the rapid growth in size and number
of position permutation loops.

Indirect experimental evidence for the position loop picture may be seen
in the behavior of the density
of liquid $^4$He on the saturation curve.
It is significant that with decreasing temperature
the density first increases approaching the $\lambda$-transition
and then decreases thereafter.
Position permutation loops
are the part of Bose-Einstein condensation
that is linked to the atomic structure of the liquid.
For liquid $^4$He above the $\lambda$-transition,
the entropy associated with position permutation loops
gives an effective short-ranged attraction to the atoms
(the weight of any one position loop increases
with decreasing distance between neighbors
due to the thermal wavelength Gaussian).
The growth in position permutation loops approaching
the $\lambda$-transition
and their decrease after it
affects the attractions between the atoms,
which is reflected in the behavior of the density.

Further experimental evidence
for the existence of position permutation loops
may be gleaned from the divergence of the heat capacity
on the high-temperature side of the $\lambda$-transition.
This cannot be due to condensation in low-lying momentum states,
since superfluidity is absent above the $\lambda$-transition.
It is instead due to energy tied up in position permutation loops
as these first increase and then decrease in number and size
with decreasing temperature.

What causes condensation to finally emerge (or re-emerge) at and below
the $\lambda$-transition?
The experimental observation that the energy and density are continuous
through the $\lambda$-transition,
and that the divergence in the heat capacity is approximately symmetrical
on each side of the transition (Lipa \emph{et al.}\ 1996)
says that the position permutation loops
begin to decline after the $\lambda$-transition,
and that they do so in a continuous manner.
The computational evidence is that the divergence in the heat capacity
is due to the growth in size of position permutation loops.
But the fact that mixed position chains
have more entropy than pure position loops of the same size
means that the mixed chains must re-emerge
at the divergent peak of the transition.
Hence position permutation loops
are increasingly converted into mixed permutation chains
on the far side of the $\lambda$-transition.
These have a condensed boson at the head,
and such condensed bosons also participate in pure momentum loops
independent of their tail.
At the peak of the $\lambda$-transition,
the large position permutation loops
as well as the  tails of the mixed chains block further growth
and the system becomes saturated.
As the temperature is decreased,
the occupation entropy of the condensed bosons increases,
which makes it favorable for long chains to be cut into smaller chains,
and the relative number of condensed bosons to increase.
The increase in condensed bosons
in highly occupied low lying momentum states
reduces the heat capacity.
This description of the far side of the $\lambda$-transition
is consistent with the number of condensed bosons increasing from zero,
and with each one converting
the position permutation loops it had been part of to a mixed chain.
The close structural relationship between pure position loops
and mixed position chains
provides a relatively smooth path
from the near to the far side of the $\lambda$-transition.

Of course it is an overly simplistic
consequence of the binary picture of condensation
that there is a single occupation entropy
applicable to every condensed boson.
A more detailed analysis gives a bespoke entropy
that depends upon the actual occupation
of the momentum state in which the condensed boson currently resides.
In this case it is the condensed bosons
that are in the most highly occupied momentum states
that are the first to be favored by the entropy of internal permutations
within their state over that of the position loops
that they would otherwise participate in.
To see this note that the near side of the $\lambda$-transition is dominated
by position permutation loops with no condensed bosons,
and the far side sees the emergence of mixed chains and condensed bosons.
The temperature difference between these two cases is a fraction of a Kelvin.
This means that the position loop entropy per boson
is almost the same as the mixed chain loop entropy per boson.
In these circumstances it is the condensed bosons
in the most highly occupied low lying momentum states
that first tip the balance.
On the far side of the $\lambda$-transition
the re-emergence of condensed bosons is continuous
rather than discontinuous,
which is to say that although the number of condensed bosons
must be macroscopic
(otherwise the temperature gradient of the energy,
the density, and the shear viscosity would be unaffected),
the fraction of condensed bosons must begin from zero
at the $\lambda$-transition itself.
This more nuanced picture gives a condensation transition
that begins with zero fraction of condensed bosons
in highly occupied states
in a more credible way than the binary all-or-nothing picture.

This can also be reconciled with the apparent discontinuity in superfluidity
at the $\lambda$-transition
by noting that it is these condensed bosons
in the most highly occupied momentum states
that selectively and most efficiently contribute to superfluid flow
in confined geometries.
The volume of these flows is generally
many times smaller than the volume of the system.
Also, the driving forces for these flows probably
selects condensed bosons in the most highly occupied momentum states.

It must be emphasized that it is not condensed bosons \emph{per se}
that are excluded from position permutation loops
and that are responsible for superfluidity.
Rather it is condensed bosons that are participating
in permutations within their own momentum state that are excluded,
and it is only when the state is highly occupied
that exclusion is favorable,
and that the viscosity is significantly reduced.

The above discussion of Bose-Einstein condensation
and the $\lambda$-transition
barely touches on the relationship between
Bose-Einstein condensation and superfluidity.
That is, at the molecular level how is it that condensed bosons
give rise to flow without viscosity?
Recent analysis and quantum molecular dynamics
computer simulations have provided
an answer to this important question (Attard 2025b).
It turns out that the rate of change of momentum
for a condensed boson in response to an applied force
is inversely proportional to the occupancy of its momentum state,
which for highly occupied states
is significantly less than that given by Newton's second law of motion.
This effect is due to the internal permutations
of the bosons within the momentum state,
and only condensed bosons involved is such momentum permutation loops
have the reduced rate of change of momentum.
Since non-equilibrium thermodynamics shows that the shear viscosity
is the time correlation function involving the time rate of change
of the first momentum moment (Attard 2012),
this explains the reduction or absence of viscosity in superfluid flow.
Condensed bosons in more highly occupied momentum states
are likely selectively involved in superfluid flow,
which explains the efficacy of this mechanism.

This dynamic mechanism is consistent
with the competition between position and momentum loops
that was discussed above as determining the $\lambda$-transition.
For a boson in a position loop,
the Gaussian thermal wavelength form for the loop bond weight
emerges after averaging over the momentum.
A condensed boson permuting with others
in a highly occupied momentum state
spends a larger amount of time with unchanged momentum
than an uncondensed boson, or a condensed boson in a few-occupied state.
In other words the conditions for the formation of a position loop
with real, positive weight
occur more frequently for an uncondensed than for a condensed boson.

%
\section{Analysis}
\setcounter{equation}{0} \setcounter{subsubsection}{0}
\renewcommand{\theequation}{\arabic{section}.\arabic{equation}}
%

\subsection{Condensed and Uncondensed Bosons}

I now give some quantitative numerical analysis
for interacting bosons
that shows the suppression of condensation above the $\lambda$-transition.
I work mainly with the binary division
of condensed and uncondensed bosons.

Consider a subsystem of $N$ interacting bosons in a cubic volume $V=L^3$
able to exchange energy with a reservoir of temperature $T$.
These bosons are divided into disjoint sets
of $N_0$ condensed bosons and $N_*$ uncondensed bosons,
with $N = N_0 + N_*$.
It would be more accurate to say that
the bosons in the set $N_*$ participate in position permutation loops,
and bosons in the set $N_0$ participate in momentum permutation loops.
The latter do not necessarily
belong to highly occupied momentum states,
although at low temperatures I expect most to do so.
I use  $N_0$
as a constraint with which to minimize the free energy,
to which there are four contributions:
the analytic quantum free energy
for $N_0$ non-interacting bosons,
the analytic classical free energy for $N_*$ non-interacting atoms,
the position loop grand potential for $N_*$ interacting bosons,
and the classical configurational integral for $N$ interacting atoms.

At high temperatures,
there are no position permutation loops,
nor are there multiply occupied momentum states.
This is the classical regime,
and the optimum numbers of the two types of bosons
are equal, $\overline N_0 = \overline N_*$, $T \gg T_\lambda$.
In this regime it is a little misleading
to call the bosons in the set $N_0$  `condensed'.

I intend to show that the optimum number of condensed bosons
just above the $\lambda$-transition vanishes, $\overline N_{0} = 0$.
Below the $\lambda$-transition temperature
I expect condensation,
$\overline N_{0} \agt \overline N_{*}$,
but I do not actually show this to be the case.
Because the momentum state and the kinetic energy of the condensed bosons
is independent of their position,
their distribution amongst the momentum states
is that of non-interacting bosons.
For the appropriate fugacity $z$,
the analytic ideal boson expression for the momentum state occupancy,
$\overline N_{\bf a}^\mathrm{id}(z)$,
satisfies $\sum_{\bf a} \overline N_{\bf a}^\mathrm{id}(z) = N_0$.
This same fugacity allows the  number of `proper' condensed bosons
to be determined,
eg.\ the bosons in multiply occupied momentum states,
$\overline N_{\bf a}^\mathrm{id}(z) > 1$.
It is emphasized that although the momentum distribution amongst
the bosons excluded from position permutation loops is ideal,
interactions between all bosons are still accounted for
in the position configuration integral, as will be seen.

\subsection{Partition Function in Phase Space}

The phase space probability density,
which was derived in earlier work (Attard 2018,  2021, 2025a Ch.~3),
is slightly modified here.
Whereas the conventional view of Bose-Einstein condensation
holds that condensation is into the ground energy state
(Le Bellac \emph{et al.}\ 2004, F. London 1938, Pathria 1972),
the actual evidence is that condensation
is into multiple low-lying momentum states (Attard 2025a Chs~2 and 3).
Most of the analysis in Ch.~3 of Attard (2025a)
was based on the so-called binary division approximation,
originally used by F. London (1938),
which indeed invokes the ground state as the single state for condensation.
The present formulation of the partition function
improves upon this by analyzing condensation into multiple momentum states.

The present model  uses an augmented phase space
that has an additional variable $s^N$,
with $s_j=0$ if boson $j$ is excluded from position loops
and $s_j=1$ otherwise.
It is not necessary to immediately take the continuum momentum limit,
and so initially all bosons
have momenta that are integer multiples of $\Delta_p = 2\pi \hbar/L$.
This quantization condition arises from the Hermiticity
of the momentum operator (Attard 2025b Appendix~B).
With this additional variable
the occupancy of the one-particle momentum state ${\bf a}$ is
given by
\begin{equation}
N_{\bf a}({\bf p}^{N_0})
=
\sum_{j=1}^N  \delta_{s_j,0}\,
\delta_{{\bf p}_j,{\bf a}}
=
\sum_{j \in N_0}
\delta_{{\bf p}_j,{\bf a}} ,
\end{equation}
where Kronecker-$\delta$s appear.

The formulation of quantum statistical mechanics
in classical phase space (Attard 2018, 2021)
introduces two functions of phase space,
namely the commutation functions, $\omega$,
which accounts for the non-commutativity
of the position and momentum operators,
and the symmetrization function, $\eta$,
which accounts for wave function symmetrization.

Bose-Einstein condensation is a non-local phenomenon
that is dominated by bosons beyond the range of the pair potential
(Attard 2025a Chs.~2 and 3).
Since the commutation function is a short-ranged function,
decaying asymptotically as the gradient of the pair potential,
the commutation function is neglected here.

The symmetrization function is the sum over all permutations
of the ratio of the original and the permuted momentum eigenfunction.
The terms in the sum lie on the unit circle in the complex plane
and mostly cancel due
to rapid oscillation with small changes in position or momentum.
The two types of permutations that survive are those between
bosons in the same momentum state,
and those between neighboring bosons in position space
that form a so-called position permutation loop.
I make the approximation that the symmetrization function
can be written as the product of pure permutations of each type
\begin{equation}
\eta({\bf q}^N,{\bf p}^N)
=
\eta_{*}({\bf q}^{N_{*}},{\bf p}^{N_{*}})
\,
\eta_{0}({\bf p}^{N_{0}}).
\end{equation}
By definition
an individual boson cannot participate in more than one permutation loop
at a time (Attard 2018, 2021).
Hence for this factorization to be valid
the two symmetrization functions have to be based upon disjoint sets.
Specifically, a condensed boson that contributes
to the momentum permutation loops embodied in $\eta_{0}$
cannot contribute to the position permutation loops in $\eta_{*}$.

The symmetrization function for the condensed bosons depends
only upon their momentum configuration,
\begin{equation}
\eta_{0}({\bf p}^{N_{0}})
=
\prod_{\bf a} N_{\bf a}! .
\end{equation}
This is the total number of ways of permuting condensed bosons
within the same momentum states.
It is the permutations of the condensed bosons
that give rise to the occupation entropy
that ultimately drives Bose-Einstein condensation and superfluidity.
In previous work (Attard 2025a Eq.~(3.3)),
the binary division approximation considered condensed bosons
to be those in the momentum ground state.
It also denoted them  with a subscript 0,
so that ${\bf p}_j = {\bf 0}$, $j \in N_0$,
and $\eta_0({\bf p}^{N_0})=N_0!$.
In contrast, the present analysis eshews condensation
solely into the ground state.

The symmetrization function for position permutation loops, $\eta_{*}$,
excludes condensed bosons, $N_{*} = N - N_\mathrm{0}$.
As just mentioned,
because any one boson can be in only one permutation loop at a time,
the factorization of the symmetrization function can only be valid
when the two factors are based on disjoint sets of bosons.
Because the spacing between momentum states goes to zero
in the thermodynamic limit,
the momentum quadrature can be performed
(Attard 2018, 2021),
\begin{eqnarray}
\eta_{*}({\bf q}^{N_{*}},{\bf p}^{N_{*}})
& \Rightarrow &
\overline \eta_{*}({\bf q}^{N_{*}})
\nonumber \\ & = &
\prod_{l=2}^{l_\mathrm{max}} e^{G^{(l)}({\bf q}^{N_{*}})} .
\end{eqnarray}
This is a mean field approximation,
where the position permutation loop symmetrization function
for the current configuration in phase space
is replaced by its momentum average,
$\overline \eta_{*}({\bf q}^{N_{*}})$.
This momentum-averaged position loop
symmetrization function is real and non-negative.
The total of the $l$-loop Gaussians is
\begin{equation}
G^{(l)}({\bf q}^{N_{*}})
=
\sum_{j_1,\ldots,j_l}^{N_{*}}\!\!'\;
G^{(l)}({\bf q}_{j_1},\ldots,{\bf q}_{j_l}) ,
\end{equation}
where the prime on the summation indicates
that all indeces must be different
and that only distinct loops are counted.
An individual $l$-loop Gaussian is
\begin{equation}
G^{(l)}({\bf q}_{j_1},\ldots,{\bf q}_{j_l})
=
\prod_{k=1}^l
e^{- \pi {q}_{j_k,j_{k+1}}^2 /\Lambda^2 } ,
\;\; {j_{l+1}} \equiv {j_1}.
\end{equation}
The thermal wave length is
$\Lambda \equiv \sqrt{2\pi\hbar^2 \beta/m}$.

With these the phase space weight is (Attard 2025a Eq.(3.61))
\begin{eqnarray} \label{Eq:w(qps)2}
\lefteqn{
w({\bf p}^{N},{\bf q}^{N},s^N|N,V,T)
}  \\ \nonumber
& = &
\frac{1}{N!V^{N}}
e^{-\beta {\cal K}({\bf p}^N)}
e^{-\beta U({\bf q}^N)}
\eta_{0}({\bf p}^{N_{0}})
\overline \eta_{*}({\bf q}^{N_{*}})  .
\end{eqnarray}
Here ${\cal K}$ is the kinetic energy, $U$ is the potential energy,
and $\beta \equiv 1/k_\mathrm{B}T$.
The momenta ${\bf p}^N$
are discrete with spacing $\Delta_p$.
Below I often use $\{{\bf q},{\bf p}\}$
as short-hand for $\{{\bf q}^N,{\bf p}^N\}$.

At high temperatures, where both the occupancy entropy,
$S^\mathrm{occ} = k_\mathrm{B} \ln \eta_{0}$,
and the position permutation entropy,
$S_* = k_\mathrm{B} \ln \overline \eta_{*}$,  are zero,
there is nothing to distinguish $s_j=0$ and $s_j=1$;
both occur with equal probability.
In this case any boson with $s_j=0$
is likely to be the sole occupant of its momentum state,
and it does not behave as condensed bosons behave.
This changes at lower temperatures
where,
due to condensed bosons being excluded from position permutation loops,
there is a competition
between the occupation entropy of condensed bosons,
and the position permutation entropy of uncondensed bosons.

The grand partition function is (cf.\ Attard 2025a Eq.~(3.1))
\begin{eqnarray}
\lefteqn{
\Xi^+(z,V,T)
}  \\
& = &
\sum_{N=0}^\infty \frac{z^N}{V^{N} N!}
\sum_{\bf p} \int \mathrm{d}{\bf q}\;
e^{-\beta {\cal H}({\bf q},{\bf p})}
\eta({\bf q},{\bf p}) \omega({\bf q},{\bf p})
\nonumber \\ & \approx &
\sum_{N=0}^\infty \sum_{{\bf s}=0,1} \frac{z^N}{V^{N} N!}
\nonumber \\ && \mbox{ } \times
\sum_{\bf p}
\int \mathrm{d}{\bf q}\;
e^{-\beta {\cal H}({\bf q},{\bf p})}
\overline\eta_{*}({\bf q}^{N_{*}}) \eta_{0}({\bf p}^{N_{0}})
\nonumber \\ & = &
\sum_{N_0=0}^\infty \sum_{N_*=0}^\infty
\frac{z^N}{V^{N} N!} \frac{N!}{N_0!N_*!}
\nonumber \\ && \mbox{ } \times
\sum_{{\bf p}}
\int \mathrm{d}{\bf q}\;
e^{-\beta {\cal H}({\bf q},{\bf p})}
\overline \eta_{*}({\bf q}^{N_{*}}) \eta_{0}({\bf p}^{N_{0}}) ,\nonumber
\end{eqnarray}
where the classical Hamiltonian is ${\cal H} = {\cal K} + U$.
The binomial coefficient is the number of ways of assigning labels
to the $N_0$ and $N_*$ condensed and uncondensed bosons;
it is the weight required for the specific boson assignment implicit
once the sum over ${\bf s}$ has been replaced
by the sums over number.
This result for the grand partition function
is equivalent to Attard (2025a Eq.~(3.3)),
except that the condensed bosons are here not confined
to the momentum ground state.

The momentum part of the grand potential for condensed bosons
is just ideal.
The corresponding grand potential is given by
\begin{eqnarray} \label{Eq:W0id}
e^{-\beta \Omega_0^\mathrm{id}}
& \equiv &
\sum_{N_0=0}^\infty
\frac{z^{N_0}}{N_0!} \sum_{{\bf p}^{N_0}}
e^{-\beta {\cal K}({\bf p}^{N_0})}
\eta_0({\bf p}^{N_0})
\nonumber \\ & = &
\sum_{N_0=0}^\infty
\frac{z^{N_0}}{N_0!}  \sum_{{\bf p}^{N_0}}
e^{-\beta {\cal K}({\bf p}^{N_0})}
\prod_{\bf a} N_{\bf a}!
\nonumber \\ & = &
\prod_{\bf a} \sum_{N_{\bf a}=0}^\infty
z^{N_{\bf a}} e^{-\beta N_{\bf a}a^2/2m}
\nonumber \\ & = &
\prod_{\bf a} [ 1 - z e^{-\beta a^2/2m} ]^{-1}.
\end{eqnarray}
The average number of condensed bosons is
\begin{eqnarray}
\overline N_0(z)
& = &
\frac{z \partial (-\beta \Omega_0^\mathrm{id})}{\partial z}
\nonumber \\ & = &
\sum_{\bf a} \frac{z e^{-\beta a^2/2m}}{1- z e^{-\beta a^2/2m}}
\nonumber \\ & \equiv &
\sum_{\bf a} \overline N_{\bf a} .
\end{eqnarray}
For a given (constrained) number of condensed bosons,
this determines the fugacity $z = \overline z(N_0)$ implicitly
as the one that satisfies $\overline N_0(z) = N_0$.

The ideal Helmholtz free energy for the condensed bosons is
$F_0^\mathrm{id}(N_0,V,T)
= \Omega_0^\mathrm{id}(\overline z,V,T)
+ \overline\mu N_0$,
with $z=e^{\beta \mu}$.
Its number derivative is  (Attard 2002 Ch.~3)
\begin{equation}
\frac{\partial F_0^\mathrm{id}(N_0,V,T)
}{ \partial N_0 }
=
\overline \mu(N_0,V,T) .
\end{equation}

The number of proper condensed bosons
(ie.\ those in multiply occupied momentum states)
for a given number of condensed bosons
could be taken to be
\begin{equation}
\overline N_{00}(N_0) =
\sum_{\bf a} \overline N_{\bf a} \, \Theta(\overline N_{\bf a}-1) .
\end{equation}
Here the unit Heaviside step function appears,
and $z = \overline z(N_0)$.
Obviously $\overline N_{\bf a}(z,V,T) $ is the average occupancy
of the momentum state ${\bf a}$ for the current number of
condensed bosons $N_0$.
This somewhat arbitrarily sets an occupancy of unity as the beginning
of proper condensation;
a larger value would likely be more useful.

A variational procedure to determine the number of condensed bosons
is required.
The constrained free energy
can be written as (cf.\ Attard 2002 Ch.~3)
\begin{equation}
\Omega(N_0|z,N,V,T)
=
F(N_0,N_*,V,T) + [N_0+N_*] \mu.
\end{equation}
Since  $N_0+N_* = N$ is held constant,
the final term can be neglected.
That is, the Helmholtz free energy $F(N_0,N_*,V,T)$
must be minimized with respect to $N_0$ at constant $N$.

Using the above expression for the ideal  Helmholtz free energy,
the Helmholtz free energy for the interacting bosons is given by
\begin{eqnarray}
\lefteqn{
e^{-\beta F(N_0,N_*,V,T)}
} \nonumber \\
& = &
\frac{1}{V^N N_0!N_*!}
\sum_{{\bf p}}
\int \mathrm{d}{\bf q}\;
e^{-\beta {\cal H}({\bf q},{\bf p})}
\overline \eta_{*}({\bf q}^{N_{*}})  \eta_{0}{\bf p}^{N_{0}})
\nonumber \\ & = &
\frac{e^{-\beta F_0^\mathrm{id}(N_0,V,T)}
}{ V^{N_0}\Lambda^{3N_*} N_*!}
\int \mathrm{d}{\bf q}^{N}\;
e^{-\beta U({\bf q}^N)} \overline \eta_{*}({\bf q}^{N_{*}}) .
\end{eqnarray}
In the final equality the sum over the momenta
of the continuum bosons
has been carried out by integration.
The  difference between this and Attard (2025a \S3.1)
is the $F_0^\mathrm{id}$.

Attard (2025a Eq.~(3.8)) developed a loop expansion for the grand potential
based on
\begin{equation} \label{Eq:exp(eta)}
\langle \eta_{*} \rangle_\mathrm{cl}
=
e^{ \langle \stackrel{\circ}{\eta}_{*} \rangle_\mathrm{cl} } .
\end{equation}
This is a classical average.
The exponent is the sum of single position loops.
This expression is valid for large systems
in the dilute compact position loop regime.
When the loops become large,
this expression improperly counts forbidden products of intersecting loops
in a way that is not negligible even in the thermodynamic limit.

The present monomer grand potential becomes
(cf.\ Attard 2025a Eq.~(3.11))
\begin{equation}
e^{-\beta \Omega^{(1)}(z,V,T)}
=
\sum_{N_0,N_*}
\frac{z^N e^{-\beta F_0^\mathrm{id}(N_0,V,T)}
}{ N_*! \Lambda^{3N_*}V^{N_0}}
Q(N,V,T) ,
\end{equation}
where the position configuration integral is
\begin{equation}
Q(N,V,T) = \int \mathrm{d}{\bf q}^N\; e^{-\beta U({\bf q}^N)} .
\end{equation}
The exponential form gives rise to the $l$-loop grand potential,
\begin{equation}
-\beta \Omega_{*}^{(l)} = \frac{(N_{*})^l}{N^{l-1}} g^{(l)},
\;\;
g^{(l)} \equiv
\frac{1}{N} \langle G^{(l)}({\bf q}^N) \rangle_{N,V,T}^\mathrm{cl} .
\end{equation}
The average of the loop Gaussian is calculated in a classical system
of $N$ particles, so that $g^{(l)}$ does not depend upon $N_{0}$.
It is intensive.
The prefactor represents the probability that $l$ bosons
chosen randomly for each loop in the original system of $N_{0}$ condensed
and $N_{*}$ uncondensed bosons are all uncondensed.

The constrained free energy is therefore taken to be
\begin{eqnarray} \label{Eq:Free1}
\lefteqn{
F(N_0|N,V,T)
 \equiv
F(N_0,N,V,T)
}  \\
\nonumber \\ & = & \nonumber
F_0^\mathrm{id}(N_0,V,T)
+ k_\mathrm{B}T
\ln [ N_*! \Lambda^{3N_*}V^{N_0}]
\nonumber \\ && \mbox{ }
- k_\mathrm{B}T \ln  Q(N,V,T)
- k_\mathrm{B}T \sum_{l=2}^{l_\mathrm{max}}
\frac{(N_{*})^l}{N^{l-1}} g^{(l)} .\nonumber
\end{eqnarray}
The derivative at constant $N$ is
\begin{eqnarray}
\lefteqn{
\frac{\partial F(N_0|N,V,T)}{\partial N_0}
}  \\
& = &
\overline \mu
- k_\mathrm{B}T \ln \frac{ N_* \Lambda^3 }{ V }
+ k_\mathrm{B}T
\sum_{l=2}^{l_\mathrm{max}}
l \left( \frac{N_{*}}{N}\right) ^{l-1} g^{(l)} .\nonumber
\end{eqnarray}
The fugacity $z=e^{\beta \mu}$ depends
on the number of condensed bosons via the ideal boson expression.
For a stable minimum in the constrained Helmholtz free energy,
the derivative passes from negative to positive with increasing $N_0$.
The condensed bosons, $N_0$,
and the uncondensed bosons, $N_*=N-N_0$,
are counted once each in the quantum  and classical ideal terms.
The condensed bosons are excluded from the position permutation loops.

At high temperatures, $z\to 0$ and $\mu \to -\infty$.
Also $ g^{(l)} \to 0$ as $\Lambda \to 0$.
Note that at high temperatures there is no distinction
between condensed and uncondensed bosons,
and $\langle s_j \rangle = 1/2$.
The quantum ideal fugacity is expected to go over to the classical
ideal fugacity,
$z^\mathrm{qu,id}(N_0) \to z^\mathrm{cl,id}(N_0)
= N_0 \Lambda^3 /V $.
Since the loop terms are zero at high temperatures,
the constrained free energy is a minimum
when the first two terms cancel,
which occurs when $\overline N_0 = \overline N_*=N/2$.
This is the expected result.

\subsection{Alternative Constrained Free Energy}

The preceding analysis is a little unrealistic
at higher temperature since in this regime
there is no real distinction
between `condensed' bosons, $s_j=0$, and uncondensed bosons, $s_j=1$,
other than the rule that the former
are excluded from position permutation loops
and also that they be analyzed
with  quantum rather than classical formulae.
It is dissonant to say that condensed bosons
are responsible for superfluidity,
when, according to this definition,
they comprise 50\% of the system
at higher temperatures where there is no superfluidity.

An alternative approach is to focus
on the lowest momentum states, ${\bf a}$,
whose kinetic energy is small compared to the thermal energy,
$\beta a^2/2m \ll 1$.
These are the states that are most likely multiply occupied
and therefore they fit better  the definition of condensed.
We can use the total number of bosons in these states,
which we denote $N_{0}'$, as the constraint.
Whereas the occupancy of the ground state is an intensive variable,
the number of bosons in states with negligible kinetic energy,
the present $N_{0}'$,
is an extensive variable and
it is therefore macroscopic (Attard 2025a \S2.5).

The constrained Helmholtz free energy
(after integrating over the remaining momenta)
is
\begin{eqnarray} \label{Eq:Free2}
\beta F(N_{0}'
\lefteqn{
|N,V,T)
=
\ln [ N_*! \Lambda^{3N_*}V^{N_{0}'}]
} \nonumber \\
&&
-  \ln  Q(N,V,T)
-  \sum_{l=2}^{l_\mathrm{max}}
\frac{N_*^l}{N^{l-1}} g^{(l)} .
\end{eqnarray}
where $N_* = N-N_{0}'$ is the number of uncondensed bosons.
As above,
the occupation entropy, $\eta_0 = \prod_{\bf a} N_{\bf a}!/N_{0}'!$,
cancels in transforming to the occupation picture
for the condensed bosons.
There is no kinetic energy contribution because it is negligible
compared to the thermal energy for all the condensed bosons.
There is no sum over momentum states for the condensed bosons
because their occupancies are specified in the constraint $N_{0}'$
(but see below).
The derivative at constant $N$ is
\begin{eqnarray}
\lefteqn{
\left( \frac{\partial \beta F(N_{0}'|N,V,T)}{\partial N_{0}'} \right)_N
} \hspace{0.4cm} \nonumber \\
& = &
-  \ln \frac{ N_* \Lambda^3 }{ V }
+
\sum_{l=2}^{l_\mathrm{max}} l \left( \frac{N_*}{N}\right) ^{l-1} g^{(l)} .
\end{eqnarray}
This passes from negative to positive with increasing $N_{0}'$
at a stable minimum in the free energy.

The ideal contribution from the condensed bosons
has been neglected in these.
If the $N_{0}'$ condensed bosons occupy $M_0$ low lying momentum states,
then the extra contribution
to the constrained Helmholtz free energy $\beta F$
would be $-M_0 \ln [ 1 + N_{0}'/M_0 ]$,
with derivative at constant occupancy
being 
$-(M_0/N_{0}') \ln [ 1 + N_{0}'/M_0 ]$.
This is negligible for large occupancies,
which is the definition of a condensed boson.

\subsection{Mixed Loops}

The preceding two subsections were based on pure loops,
namely position loops composed solely of uncondensed bosons,
and momentum loops giving permutations between condensed bosons all
in the same momentum state.
The product of the respective symmetrization functions was invoked
with the justification that since the two sets of bosons were disjoint,
the pure loops could be multiplied without violating  the condition
that a boson could belong to only one loop at a time.

However the exponential ansatz that gives rise to the loop grand potential
for the position permutation loops
already fails to enforce strictly this condition
since it implicitly allows the product of position loops
containing the same uncondensed boson.
It is believed that such overlaps are negligible
in the thermodynamic limit,
although this is likely to  be only the case
in the dilute compact loop regime.

This is one reason to consider mixed position permutation loops
containing condensed and uncondensed bosons.
A second reason is that
there appears to be missing from the above analysis
any physical mechanism to drive the re-emergence of condensed bosons
on the far side of the $\lambda$-transition.
This might be called the Lazarus transition,
because the condensed bosons that were dead on the near side
suddenly spring back to life at the peak of the $\lambda$-transition.

It is fairly clear that condensed bosons excluded from
position loops composed purely of uncondensed bosons
block and disrupt such loops.
This leads to the suppression of condensation
during the growth of position loops
on the near side of the $\lambda$-transition.
But beyond the general increase in occupancy of low-lying momentum states
with decreasing temperature there is nothing in the pure loop analysis
that specifically drives condensation once position permutation loops
have begun to form.
It is arguable that mixed loops would provide such a mechanism
since they directly link condensed bosons to the growing loops
of which they are part.

Attard (2025a \S3.2) previously analyzed mixed loops.
Much of that analysis is reproduced here
but there is one significant difference.
Attard (2025a \S3.2) argued that it was necessary to respect
the orthogonality of momentum eigenfunctions
by subtracting the disconnected contributions
to the averages of the mixed loops,
so that the multi-particle density
was replaced by the total correlation function,
giving so-called corrected (or connected) averages.
The current judgement of the author
is that such arguments are spurious,
and in what follows the ordinary average
using the usual (disconnected) multi-particle density is invoked.

Attard (2025a \S3.2.3) analyzed dressed condensed bosons
based upon an $l$-chain,
which has $l-1$ uncondensed bosons forming the tail,
and a condensed boson, labeled $j_l$, forming the head.
The condensed bosons are defined as in the preceding subsection,
namely there are $ N_{0}'$ of them, each in a low-lying momentum state,
${\bf p}_{j_l} \approx {\bf 0}$.
For this reason
chains are localized but form an open loop, $q_{j_1,j_l} \gg \Lambda$.
Non-local loops of chains may be formed head-to-tail.
There are $N_{0}'!$ ways of forming products
and concatenations of these chains.
This assumes that they have no uncondensed bosons in common,
which will be valid in the dilute condensation, small chain regime.

The $l$-chain Gaussian
for an individual condensed boson $j$ is
(cf.\ Attard 2025a Eq.~(3.36))
\begin{equation}
\tilde g^{(l)}_j({\bf q}_j;{\bf q}^{N_*})
=
\sum_{j_1,\ldots,j_{l-1}}^N \hspace{-0.3cm}'\hspace{0.2cm}
\prod_{k=1}^{l-1} e^{-\pi q_{j_k,j_{k+1}}^2/\Lambda^2},
\;\;
j_{l} \equiv j ,
\end{equation}
and $ \tilde g_j({\bf q}_j;{\bf q}^{N_*})
= \sum_{l=2}^{l_\mathrm{max}} \tilde g^{(l)}_j({\bf q}_j;{\bf q}^{N_*})$.
This can be justified by averaging the uncondensed bosons' momenta
over a contour in the complex plane,
together with the fact that the head boson
is in a low-lying momentum state,
${\bf p}_{j_l} \approx {\bf 0}$.

Actually,
just as for loops,
it is not necessary to distinguish condensed and uncondensed bosons
as the classical average in an $N,V,T$ system will soon be taken.
What distinguishes a chain from a loop
is that there is no $q_{j_l,j_1}$ bond.
Whereas loops are closed, chains are open,
they are satisfied by many more position configurations,
and they have greater weight,
particularly so as $l$ increases.

The symmetrization function is taken to be
the product of pure loops and mixed chains,
\begin{eqnarray}
\eta({\bf p},{\bf q})
& = &
\eta_*({\bf q}^{N_*}) \, \eta_0({\bf p}^{N_{0}'})
\prod_{j \in N_{0}'} [ 1 +  \tilde g_j({\bf q}_j;{\bf q}^{N_*})]
\nonumber \\ & \Rightarrow &
\eta_*({\bf q}^{N_*}) \, N_{0}'!\, [ 1 +  \tilde g ]^{N_{0}'} .
\end{eqnarray}
The term unity represents the monomer when the condensed boson
is not part of any mixed chain.
This is a mean field approximation
with $\tilde g $ (no argument)
being the classical average of the intensive chain Gaussian
(see next).

The mixed $l$-chains require the Gaussian
\begin{equation}
\tilde G^{(l)}({\bf q}^N)
=
\sum_{j_1,\ldots,j_l}^N \hspace{-0.2cm}'\hspace{0.1cm}
\prod_{k=1}^{l-1} e^{-\pi q_{j_k,j_{k+1}}^2/\Lambda^2} .
\end{equation}
From this comes the intensive form
based on the classical average,
which is independent of the division into condensed and uncondensed bosons,
\begin{equation} \label{Eq:tg(l)}
\tilde g^{(l)} \equiv
\frac{1}{N}
\langle \tilde G^{(l)}({\bf q}^N) \rangle^\mathrm{cl}_{N,V,T} .
\end{equation}
Unlike Attard (2025a \S3.2.3),
it is an ordinary classical average that appears here.
Finally, the sum total of loop Gaussians
that does depend on the number of condensed bosons is
\begin{equation}
\tilde g \equiv
\sum_{l=2}^{l_\mathrm{max}} \frac{N_{0}'N_*^{l-1}}{N^{l}}\tilde g^{(l)} .
\end{equation}
The chain weight $\tilde g $  is intensive.
The prefactor for the $l$-loop Gaussian
is the probability that of $l$ bosons randomly chosen
to form a chain, the head is a condensed boson
and the tail is composed of uncondensed bosons,
of which there are $N_* = N-N_{0}'$.

The weight of mixed loops
is defined on average rather than for a configuration
in phase space.
Multiplying the average chain weights
is valid in the small dilute loop regime.
The intensive Gaussian chain weight, $1+\tilde g$,
is like an extra fugacity for condensed bosons.
This result for the symmetrization function
links the growth of position loops to condensation,
which links superfluidity to the $\lambda$-transition.
It also suggests the mechanism by which
 the suppression of condensation on the near side
of the $\lambda$-transition is overcome.

The constrained Helmholtz free energy
(after integrating over the remaining momenta)
is
\begin{eqnarray} \label{Eq:Free3}
\lefteqn{
\beta F(N_{0}'|N,V,T)
} \nonumber \\
& = &
\ln [ N_*! \Lambda^{3N_*}V^{N_{0}'}]
-  \ln  Q(N,V,T)
\nonumber \\ && \mbox{ }
- N_{0}' \ln  [1 + \tilde g]
-  \sum_{l=2}^{l_\mathrm{max}}
\frac{N_*^l}{N^{l-1}} g^{(l)} ,
\end{eqnarray}
with $N=N_{0}'+N_*$.
As in the preceding subsection,
we have neglected
the ideal contribution from the condensed bosons.

The main difference between the loop and chain contributions
is the factor of $N_{0}'$ in the series for $\tilde g$.
Whereas increasing $N_{0}'$ only increases the loop free energy
(because it decreases $N_*$),
it can decrease the chain free energy
because of the part proportional to $N_{0}'$.
One can see that if the fraction of condensed bosons
is small and the fraction of uncondensed bosons is large,
and $\tilde g^{(l)}$ is constant or increasing with $l$,
then the sum over chains will be dominated by large chains,
and in fact the chain free energy will be reduced by increasing
the fraction of condensed bosons.
This is exactly what is required to bring back condensation
at and below the $\lambda$-transition.

The derivative at constant $N$ is
\begin{eqnarray}
 \lefteqn{
\left(
\frac{\partial \beta F(N_{0}'|N,V,T)}{\partial N_{0}'} \right)_N
}\\
& = &
-  \ln \frac{ N_* \Lambda^3 }{ V }
+
\sum_{l=2}^{l_\mathrm{max}} l \left( \frac{N_*}{N}\right)^{l-1} g^{(l)}
\nonumber \\ && \mbox{ }
-  \frac{\tilde g}{1+\tilde g} - \ln [1+\tilde g]
\nonumber \\ &  & \mbox{ }
+ \frac{1}{1+\tilde g} \left( \frac{N_{0}'}{N}\right)^2
\sum_{l=2}^{l_\mathrm{max}}
(l-1)  \left( \frac{N_*}{N}\right)^{l-2} \tilde g^{(l)} .\nonumber
\end{eqnarray}
For small $N_{0}'/N$,
the first two chain terms,
which are linear in $N_{0}'/N$,
dominate the chain series with the quadratic prefactor.
Since $\tilde g > 0$ these give a negative force
that drives an increase in the number of condensed bosons,
as was discussed above.

\section{Results}

The Monte Carlo algorithm
used to simulate Lennard-Jones $^4$He
and to obtain the loop Gaussians and the heat capacity
has been described elsewhere (Attard 2021 \S5.4.2).
Briefly,
the Lennard-Jones pair potential is
$u(r) = 4\varepsilon[(\sigma/r)^{12} - (\sigma/r)^{6}]$.
For $^4$He,
the accepted Lennard-Jones parameters are
$\varepsilon_\mathrm{He} = 10.22 k_\mathrm{B} $\,J
and $\sigma_\mathrm{He} = 0.255 6$\,nm (van Sciver 2012).
Using instead $\varepsilon = 4.8 k_\mathrm{B} $\,J
and $\sigma = 0.36$\,nm
would bring the following results into line with the measured values
for the $\lambda$-transition temperature
and for the saturated liquid density.
The Lennard-Jones pair potential
was set to zero beyond $R_\mathrm{cut}=3.5\sigma$,
and periodic boundary conditions and the nearest neighbor convention
were used.
The number of atoms in the simulations was $N=5,000$.

The Lennard-Jones $^4$He liquid saturation density at each temperature
was obtained from classical simulations of a liquid drop
in the center of the system in equilibrium with its own vapor.
Results for the heat capacity etc.\
taken over the central region of this inhomogeneous system
were in reasonable agreement with those obtained
in a separate simulation of a homogeneous system at the liquid density.
The Lennard-Jones $^4$He liquid saturation density
is about double the measured liquid saturation density of $^4$He
when using the standard value of $\sigma_\mathrm{He}$.
This is due to the approximate nature of the Lennard-Jones pair potential,
the neglect of many body potentials,
the leading one of which is the Axilrod-Teller
triple dipole potential, which is short-ranged and mainly repulsive,
and the neglect of the commutation function,
which is also short-ranged and predominantly repulsive
(Attard 2018, 2021).
Increasing $\sigma_\mathrm{He}$ by about a factor of 1.4
compensates for the neglect of these repulsive contributions
and increases the simulated saturation density
to the measured value in SI units.
It was judged best to perform the simulations
using the Lennard-Jones $^4$He liquid saturation density.

\begin{table}[tb]
\caption{ \label{Tab:Sat-gl}
Loop Gaussians for saturated Lennard-Jones $^4$He.
From classical Monte Carlo simulations with loop pruning algorithm
with the statistical error being on the order of 0.1\%
(Attard 2021).
}
\begin{center}
\begin{tabular}{c c c c c c c c}
\hline\noalign{\smallskip}
$T^*$ & $\rho^*$ & $\Lambda/\sigma$ &
 $g^{(2)}$ &  $g^{(3)}$ &  $g^{(4)}$ &  $g^{(5)}$ \\
\hline \rule{0cm}{0.4cm}%
0.35$^a$ & 1.0214 & 1.8051 & 0.559 & 0.827 & 1.57 & 3.97  \\
0.40$^a$ & 1.0190 & 1.6886 & 0.385 & 0.425 & 0.611 & 1.17 \\
0.45$^a$ & 1.0117 & 1.5920 & 0.265 & 0.223 & 0.246 & 0.366 \\
0.50$^a$ & 0.9433 & 1.5103 & 0.167 & 0.0963 & 0.0750 & 0.0763 \\
0.55$^a$ & 0.9049 & 1.4400 & 0.115 & 0.0505 & 0.0306 & 0.0242 \\
0.50$^b$ & 0.9331 & 1.5103 & 0.177 & 0.105 & 0.0869 & 0.0886 \\
0.55$^b$ & 0.9049 & 1.4400 & 0.119 & 0.0526 & 0.0333 & 0.0260 \\
0.60$^b$ & 0.8872 & 1.3787 & 0.0820 & 0.0278 & 0.0137 & 0.00854 \\
0.65$^b$ & 0.8678 & 1.3246 & 0.0565 & 0.0146 & 0.00570 & 0.00277\\
0.70$^b$ & 0.8470 & 1.2764 & 0.0389 & 0.00768 &0.00236 & 0.1911 \\
0.75$^b$ & 0.8282 & 1.2331 & 2.70E-02 & 4.08E-03 & 9.99E-04 & 3.03E-04 \\
0.80$^b$ & 0.8023 & 1.1940 & 1.84E-02 & 2.12E-03 & 4.06E-04 & 9.86E-05 \\
0.90$^b$ & 0.7503 & 1.1257 & 8.68E-03 & 5.71E-04 & 6.77E-05 & 1.02E-05 \\
1.00$^b$ & 0.7009 & 1.0679 & 4.17E-03 & 1.59E-04 & 1.18E-05 & 1.16E-06 \\
\hline
\end{tabular} \\
$^a$Droplet $^b$Homogeneous
\end{center}
\end{table}

\begin{table}[tb]
\caption{ \label{Tab:Sat-tgl}
Chain Gaussians for saturated Lennard-Jones $^4$He.
}
\begin{center}
\begin{tabular}{c c c c c c c}
\hline\noalign{\smallskip}
$k_\mathrm{B}T/\varepsilon_\mathrm{He}$ & $\rho \sigma_\mathrm{He}^3$  &
$\tilde g^{(2)}$ & $\tilde g^{(3)}$ & $\tilde g^{(4)}$ & $\tilde g^{(5)}$ \\
\hline \rule{0cm}{0.4cm}%
0.35$^a$ & 1.0214 & 2.37E+00 & 7.55E+00 & 2.55E+01 & 9.47E+01  \\
0.40$^a$ & 1.0190 & 1.84E+00 & 4.52E+00 & 1.18E+01 & 3.43E+01 \\
0.45$^a$ & 1.0117 & 1.45E+00 & 2.79E+00 & 5.73E+00 & 1.35E+01 \\
0.50$^a$ & 0.9433 & 1.07E+00 & 1.51E+00 & 2.23E+00 & 3.67E+00 \\
0.55$^a$ & 0.9049 & 8.55E-01 & 9.50E-01 & 1.11E+00 & 1.44E+00 \\
0.50$^b$ & 0.9331 & 1.13E+00 & 1.65E+00 & 2.60E+00 & 4.27E+00 \\
0.55$^b$ & 0.9049 & 8.80E-01 & 1.00E+00 & 1.23E+00 & 1.58E+00 \\
0.60$^b$ & 0.8872 & 7.01E-01 & 6.36E-01 & 6.19E-01 & 6.33E-01 \\
0.65$^b$ & 0.8678 & 5.60E-01 & 4.04E-01 & 3.12E-01 & 2.55E-01 \\
0.70$^b$ & 0.8470 & 4.47E-01 & 2.57E-01 & 1.58E-01 & 1.04E-01 \\
0.75$^b$ & 0.8282 & 3.59E-01 & 1.65E-01 & 8.13E-02 & 4.20E-02 \\
0.80$^b$ & 0.8023 & 2.85E-01 & 1.04E-01 & 4.05E-02 & 1.67E-02 \\
0.90$^b$ & 0.7503 & 1.80E-01 & 4.10E-02 & 1.01E-02 & 2.62E-03 \\
1.00$^b$ & 0.7009 & 1.14E-01 & 1.65E-02 & 2.58E-03 & 4.37E-04 \\
\hline
\end{tabular} \\
$^a$Droplet $^b$Homogeneous
\end{center}
\end{table}

Table~\ref{Tab:Sat-gl}
shows the loop Gaussians up to $l_\mathrm{max}=5$
obtained on the Lennard-Jones saturated liquid density curve.
In general the loop Gaussians decrease with increasing temperature.
At higher temperatures they decrease with increasing loop size,
but at the lowest temperatures shown
they appear to form a divergent series.
It can be seen that there is good agreement
between the results obtained from the simulation of a droplet
and those from the homogeneous simulation
at the two overlapping temperatures.

Table~\ref{Tab:Sat-tgl}
shows the chain Gaussians, Eq.~(\ref{Eq:tg(l)}).
The statistical error is on the order of 0.1\%.
As for the loop Gaussians, these decrease with increasing temperature.
They also decrease with increasing size except at the lowest temperatures.
In general the weight of a given chain
is larger than that of the corresponding loop.

At the lowest temperatures results for the free energy
for a finite number of loop Gaussians should be treated cautiously.
That  $g^{(l)}$ or $\tilde g^{(l)}$ are flat or increasing with $l$
indicates that large loops are favored.
In the large loop regime the exponential form
for the position permutation loop symmetrization function
is likely to fail because of loop intersections.
Basically when the product of a few loops are involved,
the intersections are negligible in the thermodynamic limit.
But not so for many products.
When the $g^{(l)}$ and  $\tilde g^{(l)}$
are less than unity, then only  terms with few products contribute
to the series expansion that gives the exponential.
When the loop or chain Gaussians are larger than unity,
then individual terms comprise so many products
that intersections are no longer negligible,
even in the thermodynamic limit.
For loops, the case $T^*=0.50$ is probably just acceptable,
and for chains $T^*=0.60$ is probably almost acceptable.

\begin{figure}[t]
\centerline{ \resizebox{8cm}{!}{ \includegraphics*{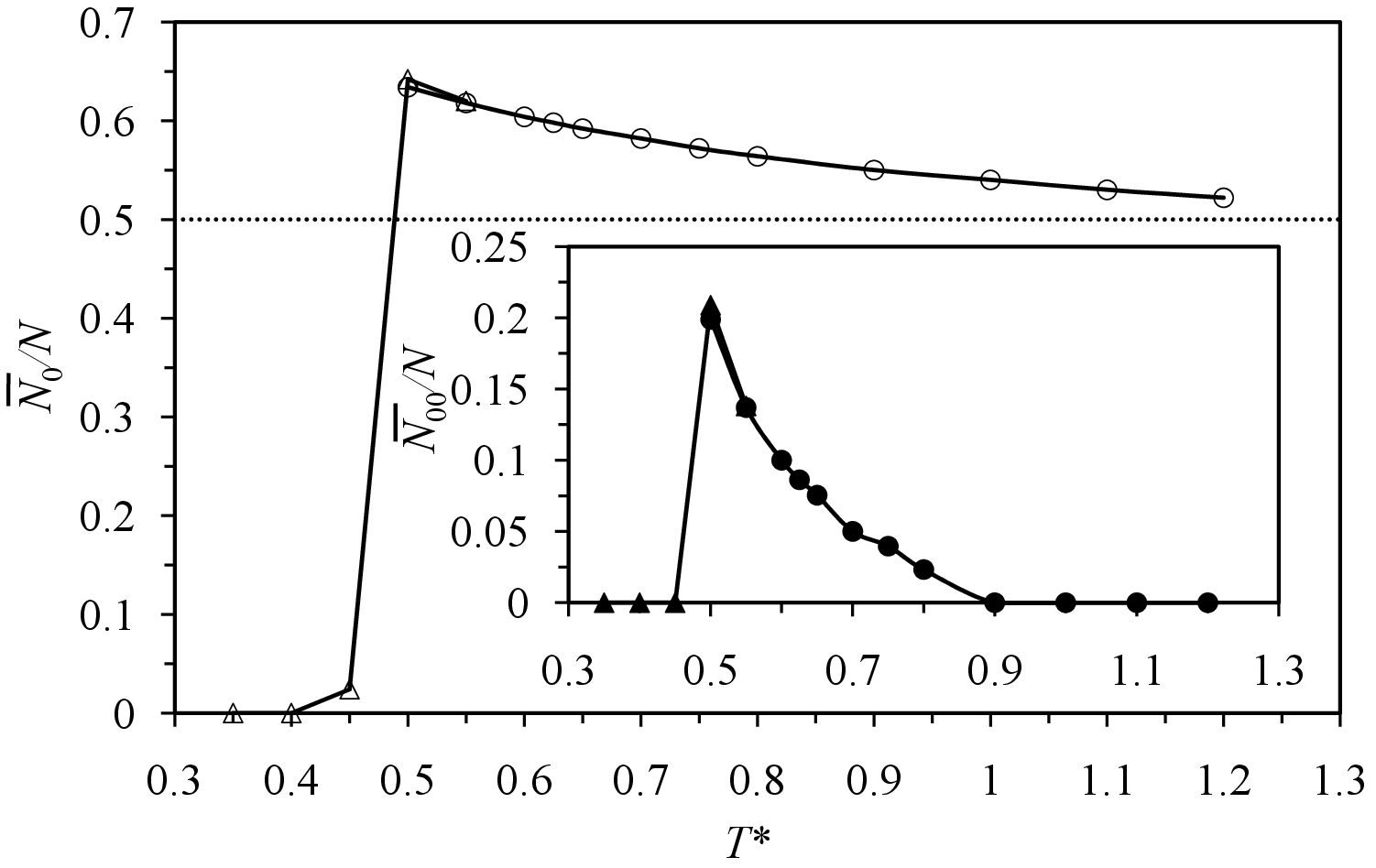} } }
\caption{\label{Fig:N0bar}
Most likely fraction of excluded Lennard-Jones $^4$He atoms,
Eq.~(\ref{Eq:Free1}).
The circles are from homogeneous simulations
and the triangles from simulations of a droplet.
The dotted line
and the lines connecting symbols are eye guides.
\textbf{Inset.}
The most likely fraction of Lennard-Jones $^4$He atoms
in momentum states with $\overline N_{\bf a} > 1$.
}
\end{figure}

Figure~\ref{Fig:N0bar} shows the most likely number of atoms
that are excluded from the position permutation loops
along the Lennard-Jones saturation curve.
This was obtained by minimizing the free energy,
Eq.~(\ref{Eq:Free1}),
using the pure position loop Gaussians in Table~\ref{Tab:Sat-gl}.
At high temperatures $\overline N_0 = N/2$, as expected.
This says that there is no difference between the classical
and quantum formulations in this regime.
As the temperature is decreased,
the fraction of quantized (excluded) bosons rises above 50\%.
This says that the contribution from the occupation entropy
increases faster than that of position loops,
so that it is favorable to use the quantum form for the ideal contribution,
which implicitly includes the occupation entropy,
to the free energy rather than the classical ideal form.
In the high temperature regime, $T^* \agt 0.5$,
the thermal wavelength is not much bigger
than the atomic diameter, $\Lambda < 1.5\sigma$,
and so there are likely to be many atoms that are not close to any neighbor,
$q_{j_k,j_{k+1}} \agt \sigma$,
which means that they are not part of loops with significant weight.
(Recall that the bond weight is $e^{- \pi q_{j_k,j_{k+1}}^2/\Lambda^2}$.)
These monomers can become condensed bosons without losing
any position permutation loop entropy.

At $k_\mathrm{B}T/\varepsilon_\mathrm{He}= 0.45$
a sharp transition takes place
to a regime in which all bosons
are included in the position permutation loops
and there are no condensed bosons.
Presumably here almost all atoms have neighbors
within the thermal wavelength
and can form position permutation loops.
Evidently the entropy to be gained from the multiple occupancy of states
is less than that lost by disrupting position permutation loops.
The requirement for such a transition
to enable consistency with experimental observation
was discussed at length in the introduction.

It should be emphasized that
at $k_\mathrm{B}T/\varepsilon_\mathrm{He}= 0.45$
the Gaussian loop series is borderline convergent,
if not actually divergent.
This means that we are likely in the regime where
large loops form.
In this regime the exponential form for the pure position loop
symmetrization function may not be accurate
because the contribution from forbidden intersecting loops
(ie.\ loops with a boson or bosons in common)
is not negligible in the thermodynamic limit.
If the exponential form is not reliable
then the constrained free energy
will be quantitatively inaccurate.
It is probably not a coincidence
that the predicted suppression of condensation
occurs just as large loops become common.

What is perhaps a little surprising in Fig.~\ref{Fig:N0bar}
are the results at higher temperatures,
namely the increase in excluded bosons from 50\%
prior to the suppression transition.
One might wonder that perhaps this suggests
the existence of superfluidity prior to the $\lambda$-transition.
However the inset to the figure makes it clear
that only a relatively small fraction of bosons
are in states with an average ideal occupancy greater than unity.
Recall that condensed bosons are defined to be bosons
in highly occupied momentum states.
The average occupancy of the states
that give rise to the data in the inset is not so large:
at $T^*=0.50$, just before the suppression transition,
the average occupancy per momentum state
for states with $\overline N_{\bf a}^\mathrm{id}(z) > 1 $
is only 6.3.
The rate of change of momentum of a condensed boson for a given force
is reduced by a factor of the inverse of the occupancy of its state,
and hence the contribution to the absence of viscosity
in superfluid flow
is greatest for condensed bosons in the most highly occupied states
(Attard 2025b).

\begin{figure}[t]
\centerline{ \resizebox{8cm}{!}{ \includegraphics*{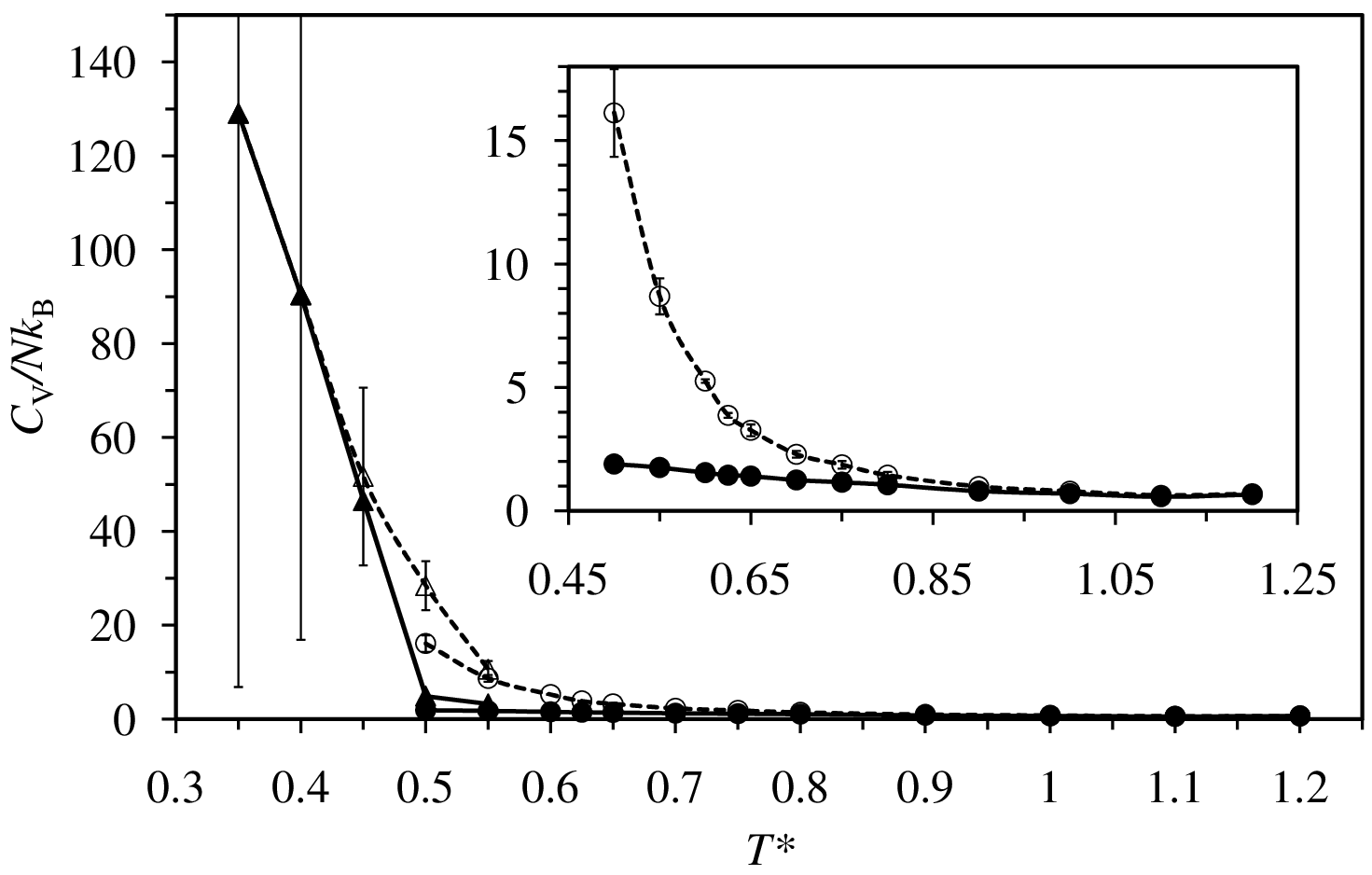} } }
\caption{\label{Fig:Cv}
The specific heat capacity of Lennard-Jones $^4$He.
The circles are from homogeneous simulations
and the triangles from simulations of a droplet.
Using Eq.~(\ref{Eq:Free1}),
the solid symbols have $\overline N_0$ atoms
excluded from the position permutation loops,
whereas the open symbols have all atoms included in the loops.
The lines connecting symbols are eye guides.
The error bars give the 95\% confidence level.
\textbf{Inset.} Magnification at higher temperatures.
}
\end{figure}

Figure~\ref{Fig:Cv} shows
the constant volume heat capacity  per atom
on the liquid saturation curve.
In the case where no atoms are excluded from the position permutation loops
(open symbols),
the heat capacity noticeably increases with decreasing temperature
for $T^* \alt 0.80$ ($T \alt 8.2$\,K using $\varepsilon_\mathrm{He}$).
However, when the optimum number of bosons
are excluded from the position permutation  loops,
the heat capacity remains rather flat
until the inclusion (suppression) transition at
$T^* = 0.45$, or $T = 4.6$\,K using $\varepsilon_\mathrm{He}$,
where it begins to increase sharply.
This is the point at which one would first observe
the approach to the $\lambda$-transition.
Between this point and the peak of the heat capacity
there are no condensed bosons and hence no superfluidity.

Experimentally,
the temperature interval
from the minimum to the peak in the measured heat capacity
on the high temperature side of the $\lambda$-transition
is about 0.4\,K
(Donnelly and  Barenghi 1998).
This shows the necessity for controlling the growth
of position permutation loops in the high temperature regime
by excluding the condensed bosons from them.
As the filled symbols in Fig.~\ref{Fig:Cv} show, 
this gives a heat capacity approaching the $\lambda$-transition
that more closely  resembles the experimental measurements
than otherwise
as it reduces the temperature range over which the heat capcity increases
and it makes the subsequent increase very sharp.
It suggests that the suppression transition should be interpreted
as the beginning of the $\lambda$-transition proper.

In the constrained free energy at lower temperatures
there is no sign of condensation re-emerging
or of a peak in the heat capacity.
Undoubtedly this is due to the simplicity of the Lennard-Jones model,
and the neglect of many body potentials and the commutation function.
Likely also the exponential form for the position loop
symmetrization function fails in the large loop regime.
Therefore the most that the data say is that the $\lambda$-transition
in Lennard-Jones $^4$He lies below $T^* = 0.45$
($T = 4.6$\,K  using $\varepsilon_\mathrm{He}$).

\begin{figure}[t]
\centerline{ \resizebox{8cm}{!}{ \includegraphics*{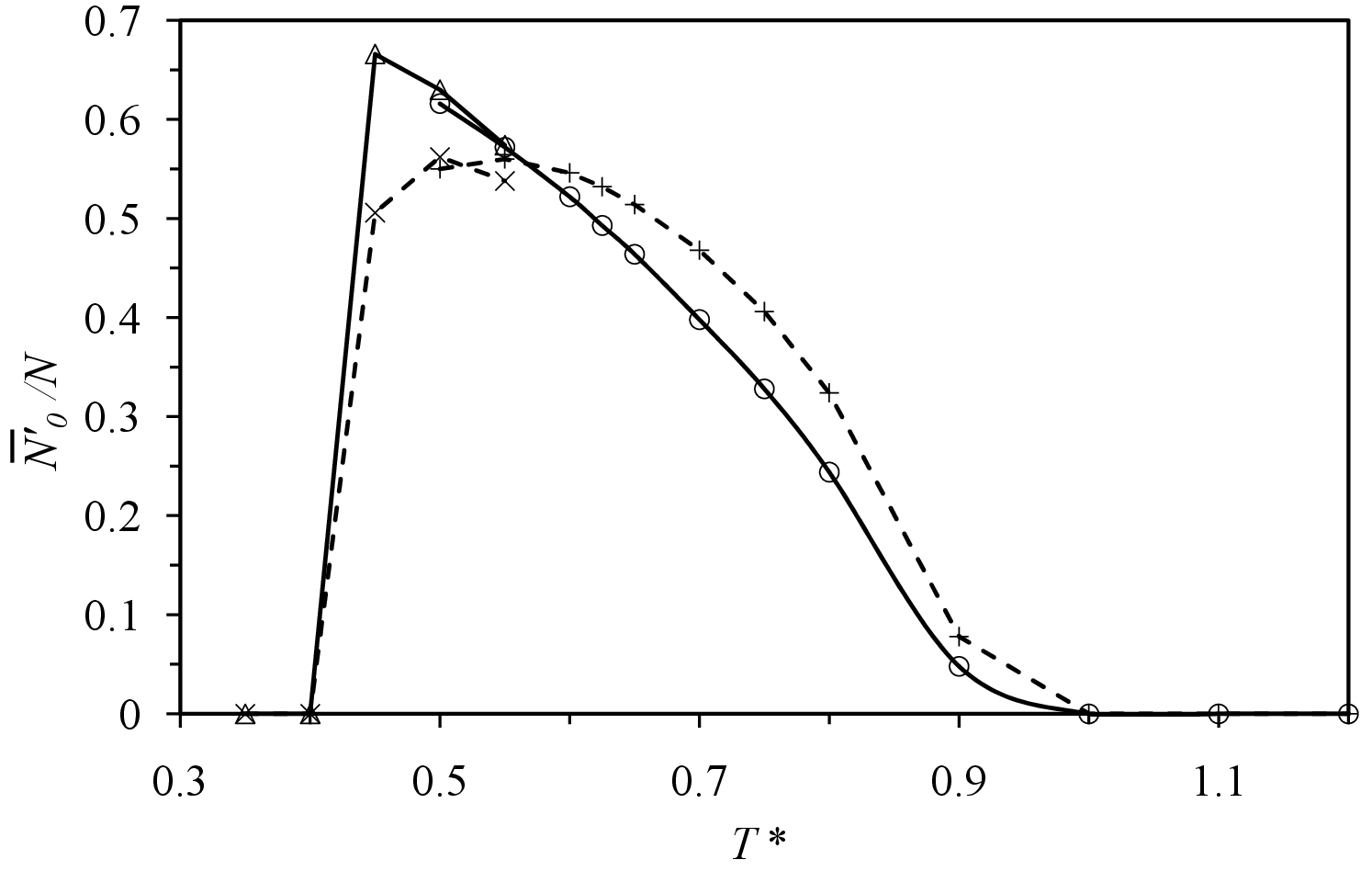} } }
\caption{\label{Fig:N0bar2}
Most likely fraction of low-lying Lennard-Jones $^4$He atoms.
Using Eq.~(\ref{Eq:Free2}) for pure loops,
the circles are from homogeneous simulations
and the triangles are from simulations of a droplet.
Using Eq.~(\ref{Eq:Free3}) including pure loops and mixed chains,
the plus symbols are from homogeneous simulations
and the times symbols are from simulations of a droplet.
The lines connecting symbols are eye guides.
}
\end{figure}

Figure~\ref{Fig:N0bar2}
shows the optimum number of condensed bosons
that minimise the alternative free energy, Eq.~(\ref{Eq:Free2}).
These are the bosons in low lying momentum states
with negligible kinetic energy
that are excluded from the position permutation loops.
It can be seen that at high temperatures there are
negligible numbers of such bosons.
In the interval $T^* \in [0.45,1.0]$
($T \in [4.6\,\mbox{K},10.2\,\mbox{K}]$ using $\varepsilon_\mathrm{He}$),
the fraction of condensed bosons increases to about 65\%
with decreasing temperature.
At about $T^*=0.4$ an inclusion transition takes place
where it has evidently become favorable to eliminate condensed bosons
and to instead include all bosons in the position permutation loops.
The behavior of $\overline N_{0}'$ shown in the figure
is similar to that shown for $\overline N_{00}$
in the inset to Fig.~\ref{Fig:N0bar},
although the peak fraction here is about three times higher.

Figure~\ref{Fig:N0bar2} shows
that the effect of including mixed chains,
Eq.~(\ref{Eq:Free3}), is surprisingly small.
The peak fraction of condensed bosons is lowered to about 56\%,
and the inclusion transition is more rounded.
The location of the inclusion transition remains unchanged
at about  4\,K.
There is no sign of condensation reemerging
due to including these mixed chains.

\begin{figure}[t]
\centerline{ \resizebox{8cm}{!}{ \includegraphics*{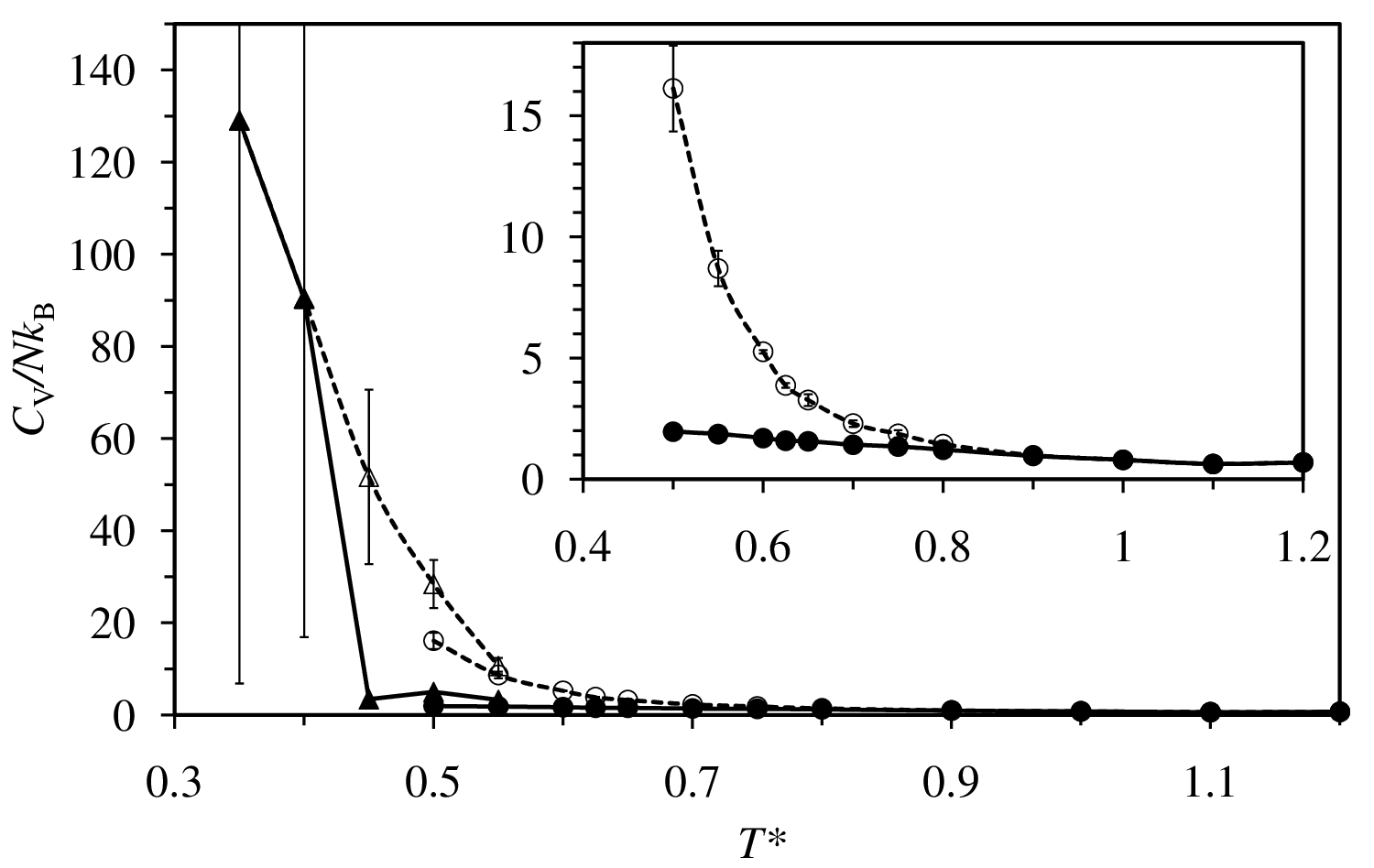} } }
\caption{\label{Fig:Cv2}
The specific heat capacity of Lennard-Jones $^4$He.
The circles are from homogeneous simulations
and the triangles from simulations of a droplet.
Using Eq.~(\ref{Eq:Free2}),
the solid symbols have $\overline N_{0}'$ atoms
excluded from the position permutation loops,
whereas the open symbols have all atoms included in the loops.
The lines connecting symbols are eye guides.
The error bars give the 95\% confidence level.
\textbf{Inset.} Magnification at higher temperatures.
}
\end{figure}

Figure~\ref{Fig:Cv2} shows the specific heat capacity
that results when $\overline N_{0}'$ bosons
are excluded from the pure position permutation loops.
Prior to the inclusion transition there is relatively little increase
in the heat capacity.
After the transition the heat capacity increases sharply.
As mentioned, the  experimentally measured width
from the peak to the minimum in the heat capacity
is about 0.4\,K (Donnelly and  Barenghi 1998).
The empty symbols in Fig.~\ref{Fig:Cv2},
which show the heat capacity if all the bosons participate
in position permutation loops,
show a significant rise over about 3\,K.
This may be misleading because because the exponential form
for the loop expansion is not reliable when large loops form,
which is likely the reason that the constrained free energy
does not show any re-emergence of condensation
or the peak of the heat capacity.
In any case the result show that
the mechanism for the divergence
of the heat capacity is the formation and growth
of position permutation loops,
which requires the suppression of condensation.

%
\section{Conclusion}
\setcounter{equation}{0} \setcounter{subsubsection}{0}
\renewcommand{\theequation}{\arabic{section}.\arabic{equation}}
%

Experimentally we may identify three temperature regimes connected
with the $\lambda$-transition in $^4$He.
The first is the high temperature regime,
from about 10\,K to 2.55\,K,
in which the heat capacity gradually decreases
with decreasing temperature to a local minimum at 2.55\,K.
The second is the near side of the $\lambda$-transition,
from 2.55\,K to 2.18\,K,
during which the heat capacity sharply increases
and diverges (integrably) to infinity.
And the third is the far side of the $\lambda$-transition,
$T < 2.18$\,K, during which the heat capacity decreases,
first rapidly and then more gradually.
It is only on the far side that  superfluidity occurs.

We may interpret these three regimes at the molecular level
on the basis of the simulation results for Lennard-Jones $^4$He.
Although the simplicity of the model fluid
and approximations such as the neglect of the commutation function
may preclude quantitative accuracy
(eg.\ the saturation density is more than twice the measured value),
nevertheless the prediction of the molecular structure ought to be
reliable as these reflect the principles of quantum statistical mechanics
and the universal attributes of Bose-Einstein condensation.

In the high temperature regime
position permutation loops begin to form,
and they grow in size and number with decreasing temperature.
However, there are also many atoms distant
from their nearest neighbors that do not participate
in position permutation loops.
Condensed atoms begin to occupy the low lying momentum states,
with their multiplicity low but increasing with decreasing temperature.
In this regime the occupation entropy per atom
is marginally greater than the position permutation loop entropy,
with on the order of 65\% of the $^4$He atoms (Fig.~\ref{Fig:N0bar})
being  excluded from the position permutation loops
(compared to 50\% in the high temperature limit).
Alternatively 55--60\% of the $^4$He atoms
are in the low lying momentum states (Fig.~\ref{Fig:N0bar2}),
which is an alternative prerequisite for condensation.
These conclusions still hold when mixed permutation chains
are included in the analysis.
The existence of condensed bosons,
which reduce the number of atoms available for  position loops,
puts a lid on their growth
and prevents any noticeable increase in the heat capacity
(cf.\ the filled symbols in the insets
to Figs~\ref{Fig:Cv} and \ref{Fig:Cv2}).
The reason that there is no superfluidity in this regime
is that the occupation of each low lying momentum state is small,
being less than about 10,
and there being less than about 20\% of the atoms
in states with an average occupancy greater than unity.

On the near side of the $\lambda$-transition,
the position permutation loop entropy per atom
exceeds the occupation entropy.
The minimum in the measured heat capacity marks the suppression transition
identified in the simulation data in the text.
On the near side of the $\lambda$-transition
it is more favorable for $^4$He atoms
to be part of position permutation  loops
than it is for them
to permute with other condensed atoms in the low-lying momentum states,
which would exclude them from the position permutation loops.
The inclusion of condensed bosons coincides
with a rapid growth in size and number
of position permutation loops, and a divergence of the heat capacity.
This conclusion is unchanged when mixed permutation chains are added,
Fig.~\ref{Fig:N0bar2}.
Because the atoms in the low-lying momentum states
do not undergo internal permutations,
even in  momentum states with high  occupancies,
there is no superfluidity.
(Actually, the results including mixed chains in Fig.~\ref{Fig:N0bar2}
indicate that there are no atoms in low-lying momentum states
in this regime,
which is a more direct way of seeing that there can be no superfluidity.)

I have not directly accessed the far side of the $\lambda$-transition
in the simulations, and so what follows is somewhat speculative.
(But see the appendix added in version 2.)
The key challenge is to explain
why the $\lambda$-transition is marked
by an infinite rather than a finite maximum heat capacity.
The results in Figs~\ref{Fig:N0bar} and \ref{Fig:N0bar2}
indicate that on the near side of the $\lambda$-transition
condensed boons are suppressed,
and the results in  Figs~\ref{Fig:Cv} and \ref{Fig:Cv2}
indicate that the heat capacity diverges,
presumably to infinity.
The results in Tables~\ref{Tab:Sat-gl} and \ref{Tab:Sat-tgl}
indicate that the two loop series are divergent in this regime.
This says that large pure position loops 
are forming on the near side.
(There can be no mixed position chains because
there are no condensed bosons to form each head.)
Since loops are closed and chains are open,
when there is no distinction between condensed and uncondensed bosons
there are many more possible chains than loops
$g^{(l)} < \tilde g^{(l)}$,
with the difference growing exponentially with $l$.
At the point of divergence of the pure position loop series,
which is presumably the point of infinite heat capacity
and the true $\lambda$-transition,
it becomes favorable for mixed chains to emerge.

The molecular mechanism for this is based on the fact
that for any given position configuration and any given permutation,
converting an uncondensed boson to a condensed boson
increases the weight of the position permutation loop
that it is part of by a factor of approximately
$e^{\pi \sigma^2/\Lambda^2}$.
This is not true for a monomer in the particular permutation,
but in this regime no atom will remain a monomer
for more than a small minority of permutations.
A similar result holds for mixed chains once they emerge,
so that long chains are favorably broken into short chains.
This appears to be the mechanism by which condensed bosons re-emerge
at the peak of the $\lambda$-transition
and by which their fraction increases with decreasing temperature
on the far side of the transition.

The experimental observation is that condensation begins from zero
at the $\lambda$-transition,
as evidenced by the continuity in density and shear viscosity,
and the discontinuity in their first temperature derivative.
The present explanation for the $\lambda$-transition
based on the emergence of chains is consistent
with that observation in so far as
the entropy due to the occupancy of individual momentum states
is a collective phenomena,
which non-linearity drives the transition.
But since occupancy is an intensive variable,
it is possible for individual momentum states to become highly occupied
for the first time at the $\lambda$-transition,
and at the same time the total number of condensed bosons
is macroscopically zero.
As the temperature is lowered on the far side of the transition,
the number of highly occupied momentum states increases
such that the  fraction of condensed bosons becomes macroscopic.

The reason that the heat capacity decreases as chains replace loops
and short chains replace long chains
is because of the increasing fraction of condensed bosons,
since the ones in low-lying momentum states
contribute little to the heat capacity.
Also the number of permutation loops and chains has probably maxed out,
because large loops and chains block others from forming.

In analyzing pure position permutation loops and mixed chains,
the bond weight is a Gaussian with thermal wavelength width.
This comes from averaging over momentum states.
For the case of uncondensed bosons,
changes in momentum state are much more frequent
than significant changes in position,
and so such averaging is valid.
For the case of condensed bosons the rate of change of momentum
is inversely proportional to the occupancy of its momentum state,
and so it remains in the same momentum state for relatively long periods.
Hence it is valid to treat condensed bosons as if they
are a separate species.

Superfluidity is due to the permutations between bosons
within highly occupied momentum states (Attard 2025b).
It should be noted that in arguing above
for the emergence of mixed chains
on the far side of the $\lambda$-transition,
in the corresponding symmetrization function,
$N_{0}'! \, [ 1 + \tilde g]^{N_{0}'}$
(the prefactor would in more precise analysis
be replaced by $\prod_{\bf a} N_{\bf a}!$),
the term unity represents the internal permutations
of the individual bosons within each momentum state
(ie.\ independent of the uncondensed bosons in the tail of each chain).
Thus the emergence of chains on the fair side of the $\lambda$-transition
is consistent with an increase in the occupation entropy
due to internal permutations within momentum states.

On the far side of the $\lambda$-transition,
the occupancies of the low lying momentum states
are larger, presumably many times larger,
than they were in the high temperature regime.
This means that the occupation entropy
is larger due to the internal permutations
within each low lying momentum state.
It also means that for these condensed atoms
the rate of change of momentum is greatly reduced,
and that the number of atoms exhibiting such a reduction
is greatly increased.
These together give rise to superfluidity.

\vspace{1cm}

\section*{References}


\begin{list}{}{\itemindent=-0.5cm \parsep=.5mm \itemsep=.5mm}

\item %
Attard P 2002
\emph{Thermodynamics and Statistical Mechanics:
Equilibrium by Entropy Maximisation}
(London: Academic)

\item 
Attard  P 2012
\emph{Non-equilibrium thermodynamics and statistical mechanics:
Foundations and applications}
(Oxford: Oxford University Press)

\item 
Attard P 2018
Quantum statistical mechanics in classical phase space. Expressions for
the multi-particle density, the average energy, and the virial pressure
arXiv:1811.00730

\item 
Attard P 2021
\emph{Quantum Statistical Mechanics in Classical Phase Space}
(Bristol: IOP Publishing)

\item 
Attard P 2025a
\emph{Understanding Bose-Einstein Condensation,
Superfluidity, and High Temperature Superconductivity}
(London: CRC Press)

\item
Attard P 2025b
The molecular nature of superfluidity: Viscosity of helium from quantum
stochastic molecular dynamics simulations over real trajectories
arXiv:2409.19036v5

\item 
Le Bellac M, Mortessagne F, and Batrouni G G 2004
\emph{Equilibrium and Non-Equilibrium Statistical Thermodynamics}
(Cambridge: Cambridge University Press)

\item 
Donnelly R J and  Barenghi C F 1998
The observed properties of liquid Helium at the saturated vapor pressure
\emph{J.\ Phys.\ Chem.\ Ref.\ Data} {\bf 27} 1217

\item 
Einstein A 1924
Quantentheorie des einatomigen idealen gases
Sitzungsberichte der Preussischen Akademie der Wissenschaften
{\bf XXII} 261 

\item 
Einstein A  1925
Quantentheorie des einatomigen idealen Gases. Zweite Abhandlung.
Sitzungsberichte der Preussischen Akademie der Wissenschaften
{\bf I} 3


\item 
Lipa J A, Swanson D R, Nissen J A, Chui T C P, and Israelsson U E 1996
Heat capacity and thermal relaxation of bulk helium
very near the lambda point
\emph{Phys.\ Rev.\ Lett.}\ {\bf 76} 944

\comment{
$C_S = A t^{-\alpha}[ 1 + D t^{1/2} + E t ] + B$,
where $t \equiv |1 -T/T_\lambda|$,
$\alpha = -0.012 850$,
$A = 5.7015$ for $T < T_\lambda$ and  $A = 6.0092$  for $T > T_\lambda$,
$B=456.28$, $D=-0.0228$, and $E=0.323$.
It is possible that the units of $A$ and $B$ and $J /(\mathrm{mol}\,K)$.
(These are gram moles).
}

\item  
London F 1938
The $\lambda$-phenomenon of liquid helium and the Bose-Einstein degeneracy
\emph{Nature} {\bf 141} 643

\item 
Pathria R K 1972
\emph{Statistical Mechanics} (Oxford: Pergamon Press)

\item 
van Sciver  S W 2012
\emph{Helium Cryogenics}
(New York: Springer 2nd edition)

\item  
Tisza L 1938
Transport phenomena in helium II
\emph{Nature} {\bf 141} 913

\end{list}



\appendix
%
\section{Refined Free Energy, Fugacity, and Chains}
\renewcommand{\theequation}{\Alph{section}.\arabic{equation}}
%

This appendix extends the text (version 1) in two ways.
First it gives a better expression
for the free energy and fugacity of the condensed bosons.
And second,
the mixed position permutation chains
are reformulated in order to resolve an inherent contradiction
in their treatment in the text.
The conclusions in the text are largely unchanged,
except that more detail and certainty are provided
for the emergence of condensation at and below the $\lambda$-transition.

\subsection{Free Energy and Fugacity}

In the text the fugacity $z$,
which is required for the condensed boson grand potential,
$\Omega_0^{\rm id}(z,V,T) = \prod_{\bf a} [ 1 -z e^{-\beta a^2/2m} ]^{-1}$,
Eq.~(\ref{Eq:W0id}), is determined implicitly
by setting $\overline N_0^{\rm id}(z) = N_0$,
where $ \overline N_0^{\rm id}(z) =
\sum_{\bf a} z e^{-\beta a^2/2m} /[ 1 -z e^{-\beta a^2/2m} ]$
is evaluated numerically
by the sum over momentum states for a small system.
Here an arguably better approach improves this in two ways.

First, the continuum approximation is made,
so that the grand-potential is given by the Bose-Einstein integral
(Pathria 1972 Appendix~D, Attard 2025 \S2.3)
\begin{equation}
\Omega_0^{\rm id}(\tilde z_0,V,T)
=
-k_{\rm B}T V \Lambda^{-3} g_{5/2}(\tilde z_0) ,
\end{equation}
and
\begin{equation}
\overline N_0^{\rm id}(\tilde z_0)
=
 V \Lambda^{-3} g_{3/2}(\tilde z_0) .
\end{equation}
For these to be meaningful the condensed boson fugacity
must be less than unity, $\tilde z_0 <1$,
as is discussed below.

The argument against the continuum approximation in the past has been
that it fails to capture the contribution from the ground state bosons
(Pathria 1972 \S7.1, Attard 2025 \S2.3).
The basis for this argument is physical rather than mathematical:
the maximum ideal boson density that results
is less than the measured density of saturated liquid $^4$He.
Therefore, it is said, the continuum represents
only the excited state bosons,
and extra ground state bosons need to be added
to make up the discrepancy,
$N_0 = N^{\rm meas}_{\rm 4He} - \overline N^{\rm id}_*(z)$.
This has the unfortunate consequence
that the fugacity becomes dependent on the system size,
which violates the fundamental thermodynamic principle
that it is  an intensive variable.
In fact, however, the problem lies not with the continuum approximation,
which is quite accurate (Attard 2025a \S2.4),
but with the ideal boson model itself.
Simply adding the ideal boson ground state to the continuum result
compensates for one approximation (the ideal boson model)
by introducing another (the added ground state).

In the present approach interactions between bosons
are included via the position configuration integral $Q(N,V,T)$,
and also via the permutation loops and chains.
Thus there is no real reason to follow the ideal gas model
and to suppose that the continuum integral only applies
to the excited state bosons and that the ground state
bosons have to be added separately.
Of course the analysis in the text is based
on treating condensed and uncondensed bosons as separate species
and optimizing their number.
The point being made here is that the grand potential
of the condensed bosons can be obtained via the continuum integral.

For the second improvement,
the fugacity of the condensed bosons is taken to be
\begin{equation}
\tilde z_0(N_0) =  f_0 \rho \Lambda^3 [ 1 + \tilde g ],
\end{equation}
where $\rho=N/V$ is the number density of the liquid
(in the results below this is taken to be
that of the  Lennard-Jones $^4$He saturated liquid),
and $f_0 = N_0/N$ is the fraction of condensed bosons,
which the chain contribution depends upon,
$\tilde g = \sum_{l=2}^{l_{\rm max}} f_0 f_*^{l-1} \tilde g^{(l)}$,
the fraction of uncondensed bosons being $f_* = 1 - f_0$.
The part $ f_0 \rho \Lambda^3 $
is the momentum contribution to the fugacity for ideal bosons,
which is augmented with the contribution from the permutation chains.
The difference with the text is that this is an explicit expression
for the fugacity $\tilde z_0(N_0)$
rather than implicit equation to be solved numerically.
Of course the consequence is that
$\overline N_0^{\rm id}(\tilde z_0) \ne N_0$.
(These do not have to be equal because the excess contribution
from the position configuration integral also contributes.)

One  merit of the present expression for the fugacity
is that at high temperatures
where the chains vanish,
$\tilde g \to 0$, it goes over to the ideal boson result.
Since in this limit the loops also vanish, $g \to 0$,
the condensed and uncondensed boson are treated identically
with the momentum contributions of both being ideal.
The interactions for both types of atom are accounted for
by the position configuration integral.

With the continuum expression for the condensed boson grand potential,
the constrained Helmholtz free energy becomes
\begin{eqnarray}
\lefteqn{
\beta F(N_{0}| N,V,T)
} \nonumber \\
& = &
- V \Lambda^{-3} g_{5/2}(\tilde z_0)  + N_0  \ln \tilde z_0
+
\ln [ N_*! \Lambda^{3N_*} V^{N_0} ]
\nonumber \\ && \mbox{ }
-  \ln  Q(N,V,T)
-  \sum_{l=2}^{l_\mathrm{max}}
\frac{N_*^l}{N^{l-1}} g^{(l)} ,
\end{eqnarray}
with $N=N_{0}+N_*$ and $z_0(N_0)<1$.
Per atom this is
\begin{eqnarray}
\lefteqn{
\beta F(N_{0}| N,V,T)/N
} \nonumber \\
& = &
\frac{-1}{\rho \Lambda^3} g_{5/2}(z_0)  + f_0 \ln z_0
+ f_* \ln [f_* \rho \Lambda^3 ] - f_*
\nonumber \\ && \mbox{ }
-  \sum_{l=2}^{l_\mathrm{max}} f_*^l g^{(l)}
+ \ln V
-  N^{-1}\ln  Q(N,V,T).
\end{eqnarray}
Only the terms that depend on $f_0$, $f_*$, and $\tilde z_0$ need to be kept.
The optimum fraction of condensed bosons is obtained by minimizing this.

Equivalently, and often more conveniently,
one can simply find the zero of the derivative at constant $N$,
\begin{eqnarray}
\lefteqn{
\left(
\frac{\partial \beta F(N_{0}|N,V,T)}{\partial N_{0}} \right)_N
}\\
& = &
\ln \tilde z_0
+ \left[ N_0 - \overline N_0^{\rm id}(\tilde z_0) \right]
\frac{\partial \tilde z_0}{\tilde z_0\partial  N_0 }
-  \ln \frac{ N_* \Lambda^3 }{ V }
\nonumber \\ && \mbox{ }
+
\sum_{l=2}^{l_\mathrm{max}} l \left( \frac{N_*}{N}\right)^{l-1} g^{(l)}.
\end{eqnarray}
One has
\begin{eqnarray}
\frac{\partial \tilde z_0}{\tilde z_0\partial  N_0 }
& = &
\frac{1}{N f_0 [1+\tilde g]}
\left\{  \rule{0cm}{0.5cm}
[1+\tilde g]
\right.  \\ \nonumber && \mbox{ } \left.
+
f_0
\sum_{l=2}^{l_{\rm max}}
\left[  f_*^{l-1} - (l-1)f_0 f_*^{l-2} \right] \tilde g^{(l)}
\right\} .
\end{eqnarray}
Since $\overline N_0^{\rm id}(\tilde z_0) \sim N_0 [1+\tilde g] $,
$N_0 \to 0$,
it follows that
$\lim_{N_0 \to 0}  [ N_0 - \overline N_0^{\rm id}(\tilde z_0) ]
{\partial z_0}/{z_0\partial  N_0 }
\to 0$, $N_0 \to 0$.

As mentioned above these expression are only meaningful for
$\tilde z_0 < 1$,
which is guaranteed for low condensed boson fractions,
specifically for $f_0=0$.
At higher fractions and lower temperatures,
$\tilde g$ becomes large and this condition is violated.
The numerical results (next) for the optimum fraction $\overline f_0$
satisfy this condition.
At those temperatures where not all fractions are valid,
$\overline f_0$ gives the lowest free energy,
usually at a boundary of the valid range (or ranges),
where the derivative does not vanish.

\begin{figure}[t]
\centerline{ \resizebox{8cm}{!}{ \includegraphics*{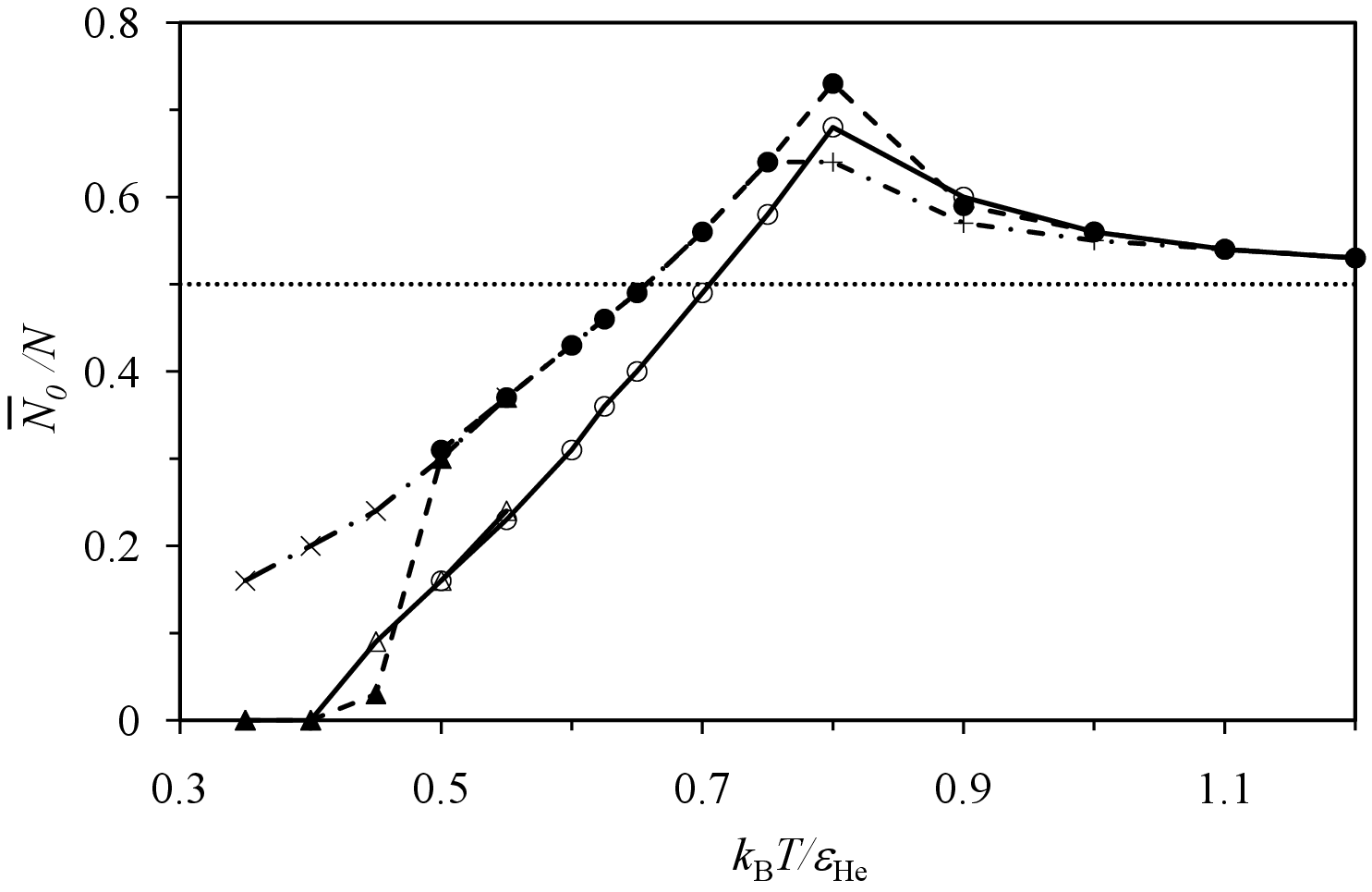} } }
\caption{\label{Fig:N0barA}
Most likely fraction of condensed Lennard-Jones $^4$He atoms.
The circles are from a homogeneous liquid (periodic boundary conditions)
and the triangles are from simulations of a droplet.
The empty symbols are for position loops and chains,
and the filled symbols are for position loops only.
The algebraic symbols are from Eq.~(\ref{Eq:Ftot3}).
The dotted line and the lines connecting the symbols are eye guides.
Note that $\varepsilon_{\rm He}/k_{\rm B} = 10.22$\,K.
}
\end{figure}

Figure~\ref{Fig:N0barA} shows the optimum fraction of condensed bosons
with (empty symbols) and without (filled symbols)
the chain contribution.
At high temperatures the loops and chains are negligible,
and the optimum faction approaches 50\%
as there is nothing to distinguish condensed from uncondensed bosons
in this regime.
Because of this the nomenclature `condensed' is a little misleading.
As the temperature is decreased,
the optimum fraction of condensed bosons first  rises slowly above 50\%,
and then abruptly begins to fall.
Compared to the results in the text,
specifically Fig.~\ref{Fig:N0bar},
the decrease in the fraction of condensed bosons
with decreasing temperature is more gradual here,
and it begins at a higher temperature.
Including the chains  gives a lower fraction of condensed bosons,
and a more gradual fall to zero condensation
toward the end of the temperature range shown.
In both cases condensation is completely suppressed
at the lowest temperatures studied.
There is quite good agreement between the simulation results
for the homogeneous liquid (circles)
and the droplet in equilibrium with its own vapor (triangles)
where their temperature range overlaps.

The first temperature
at which $\tilde z_0(N_0) > 1 $
is $T^* = 0.90$, namely at the constrained fraction $f_0=0.93$ (with chains)
or  $f_0=0.94$ (without chains).
(These are not the optimum fraction.)
This constrained fraction decreases with decreasing temperature.
Temperatures above the cusp at $T^*=0.80$ evident in Fig.~\ref{Fig:N0barA}
have the minimum in the free energy located within the domain
$\tilde z_0(N_0) < 1 $,
whereas for temperature at and below this cusp
the minimum occurs at the boundary of the valid domain.

Also shown in Fig.~\ref{Fig:N0barA}
are the results of the free energy  Eq.~(\ref{Eq:Ftot3}),
which includes the chain grand potential similarly to the loops
(see below).
The results are rather similar to the results without chains,
except that they don't jump to zero condensed boson fraction
at the lowest temperatures.

\begin{figure}[t]
\centerline{ \resizebox{8cm}{!}{ \includegraphics*{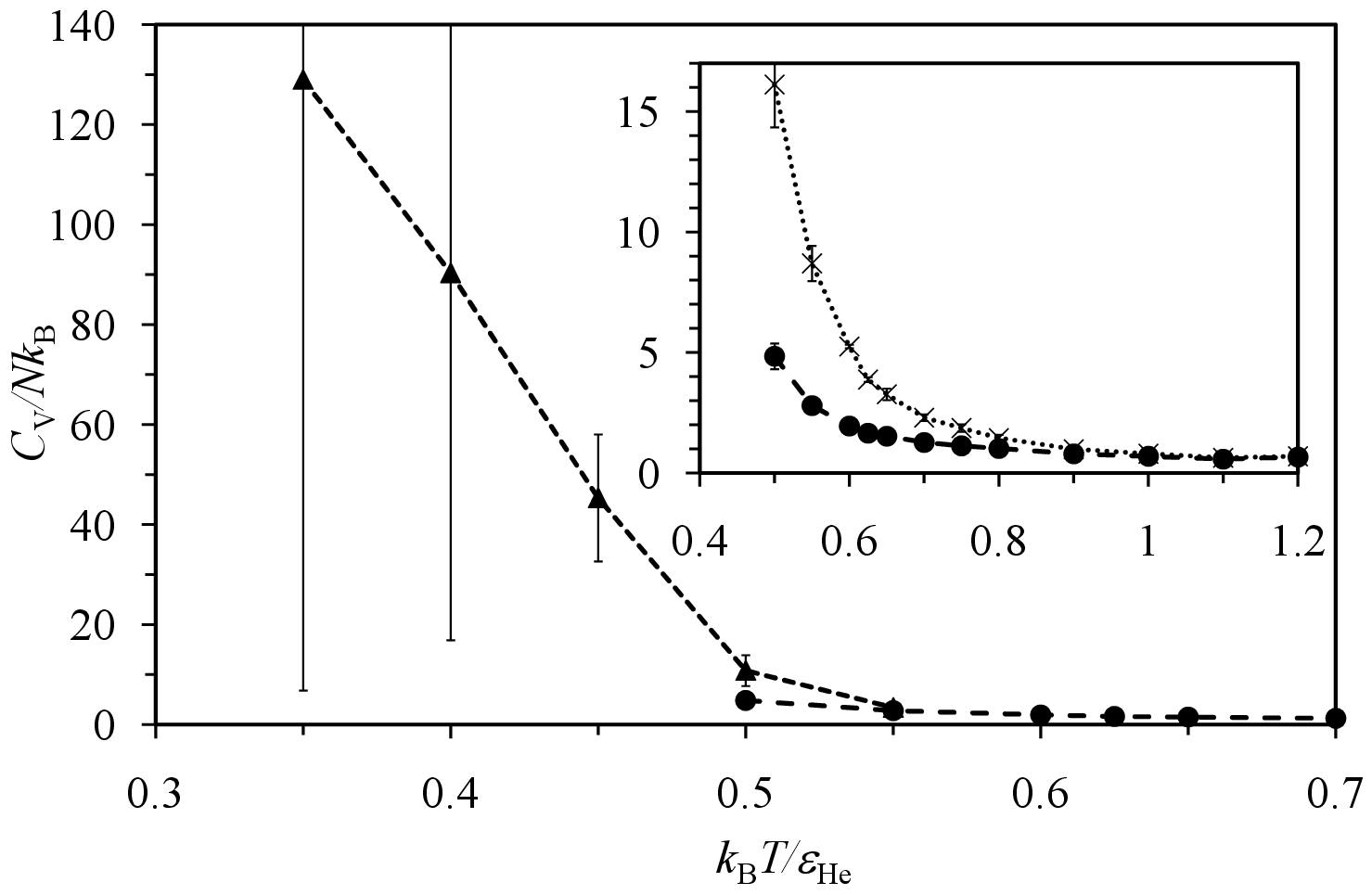} } }
\caption{\label{Fig:CvA}
The specific heat capacity of Lennard-Jones $^4$He
due to the classical contribution plus position loops only.
The circles are from a homogeneous liquid
and the triangles are from simulations of a droplet.
The filled symbols have the $\overline N_0$ condensed atoms
excluded from the position permutation loops,
whereas the crosses in the inset have all atoms
included in the position loops.
The lines connecting symbols are eye guides.
The error bars gives the 95\% confidence level.
}
\end{figure}

Figure~\ref{Fig:CvA} shows the specific heat capacity at constant volume.
No contribution from the mixed permutation chains   is included here.
There is a sharp increase in the specific heat capacity
for $T^* \alt 0.5$ ($T \alt 5.11$\,K).
This corresponds to the regime in which condensation is suppressed
and all atoms are therefore included in the position permutation loops.
This divergence in the heat capacity is due to the divergence of the
position permutation loop series.
It can be seen in the inset that the non-zero number of condensed bosons,
and their exclusion from the position permutation loops
for $T^* \agt 0.5$
significantly reduces and flattens the heat capacity
compared to if all atoms had participated in the position permutation loops.
This has the effect of sharpening the high temperature side
of the $\lambda$-transition.
Again the differences with the results in the text are quantitative
rather than qualitative.

The Lennard-Jones saturation liquid density is
about 2.7 times the measured density.
This causes the parameter $\rho \Lambda^3$
to be overestimated, as well as the peak of the pair distribution function.
These are the reasons that the heat capacity diverges
at a higher temperature in the Lennard-Jones liquid
than measured in $^4$He.

\subsection{Permutation Chain Grand Potential}

In the preceding section
the condensed bosons are `dressed' with
the mixed chains with which they are associated,
which gives rise to the effective fugacity
$\tilde z_0 = z_0^{\rm id,cl} [1+\tilde g]$.
An alternative approach is to treat the chains similarly to the loops
by defining the chain grand potential.
The justification for this
may be illustrated by the double dimer permutation,
\begin{eqnarray}
\eta^{(2,2)}({\bf \Gamma})
& = &
\sum_{j<k}^{N_0}\sum_{l,m}^{N_*}\!^{(l \ne m)}
e^{-{\bf p}_l \cdot {\bf q}_{lj}/{\rm i}\hbar}
e^{-{\bf p}_m \cdot {\bf q}_{mk}/{\rm i}\hbar}
\nonumber \\ & \approx &
\frac{1}{2} \sum_{j=1}^{N_0}\sum_{l=1}^{N_*}
e^{-{\bf p}_l \cdot {\bf q}_{lj}/{\rm i}\hbar}
 \sum_{k=1}^{N_0}\sum_{m=1}^{N_*}
e^{-{\bf p}_m \cdot {\bf q}_{mk}/{\rm i}\hbar}
\nonumber \\ & = &
\frac{1}{2}  \eta^{(2)}({\bf \Gamma})^2 .
\end{eqnarray}
Dropping the restriction $l \ne m$ gives negligible error
in the thermodynamic limit.
The factor of one half obviously represents the second term
in an infinite exponential series of products
that can be extended to all chain lengths,
just as for position permutation loops
(Attard 2018, 2021, 2025a \S7.3.3).

The corresponding Helmholtz free energy is given by
\begin{eqnarray} \label{Eq:Ftot3}
\lefteqn{
\beta F(N_0, N_*,V,T)
}  \\
& = &
\Omega_0(z_0,V,T) + N_0 \ln z_0
+
\ln [N_*! \Lambda^{3N_*} V^{-N_*} ]
\nonumber \\ && \mbox{ }
- \ln [ Q(N,V,T) /V^{N} ]
- N g(f_*)
- N \tilde g(f_0,f_*) ,\nonumber
\end{eqnarray}
where $  g(f_*) = \sum_l f_*^l g^{(l)}$
and $\tilde g(f_0,f_*)
= \sum_l f_0 f_*^{l-1} \tilde g^{(l)}$.
With $z_0 = f_0 \rho \Lambda^3$,
the derivative at constant $N$ is
\begin{eqnarray}
\lefteqn{
\frac{\beta}{N} \frac{\partial F(N_0, N_*,V,T)}{\partial f_0}
}  \\
& = &
\ln z_0
+ \left[ 1 - z_0^{-1} g_{3/2}(z_0)\right]
- \ln [f_* \rho \Lambda^{3} ]
\nonumber \\ && \mbox{ } \nonumber
+  \sum_{l=2}^{l_{\rm max}} l f_*^{l-1} g^{(l)}
-  \sum_{l=2}^{l_{\rm max}}
\left[ f_*^{l-1} - (l-1) f_{0}  f_*^{l-2} \right] \tilde g^{(l)}.
\end{eqnarray}

The results in Fig.~\ref{Fig:N0barA}
show that this alternative free energy expression
is not significantly different to that given in the preceding section.
Perhaps the most noticeable difference
is that condensation is partially but not entirely suppressed
at the lowest temperatures shown.
One should be cautious about the conclusions drawn
from these free energy expressions because of their approximate nature,
the simplicity of the Lennard-Jones model,
and the limited number of terms that are used
in the loop and chain series.

\subsection{Summary}

The author's present understanding of the $\lambda$-transition
may be summarized as:

\begin{enumerate}
\item
On the high temperature side of the $\lambda$-transition
position permutation loops form.
The links have Gaussian weight dependent upon the bond length,
$e^{- \pi q_{j_k,j_{k+1}}^2 /\Lambda^2}$,
where the thermal wavelength $\Lambda$ comes from the momentum integration
with Maxwell-Boltzmann weight, $e^{-\beta p_{j_k}^2/2m}$.
There are few condensed bosons in this regime
because there are many more accessible momentum states
than there are atoms.

\item
As the temperature is decreased,
the driving force for condensed bosons
(ie.\ those in multiply occupied low-lying momentum states) increases.
This is due (1) to the fewer available momentum states,
(2) the occupation entropy of multiply occupied states,
and (3) the chain entropy.

\item
Chains begin as loops but with a longer bond connecting
the condensed boson (the head)
with the adjacent non-condensed boson (the tail).
The longer bond arises from the greater weight given to
low-lying momentum states than the Maxwell-Boltzmann weight alone.
The average end-to-end distance of a chain decreases
with increasing momentum state of the condensed boson.

\item
Bosons in the same momentum state,
which in practice means the condensed bosons
in multiply occupied low-lying momentum states,
are permuted non-locally independently of any chain
that they can also form.
These non-local permutations and the consequent occupation entropy
give rise to superfluidity.

\item
Chains occur below the $\lambda$-transition,
as experimentally evidenced by the decrease in density
and the decrease in the specific heat capacity.
Because chains are less compact than loops
the density decreases below the $\lambda$-transition.
Because condensation leads to shorter chains and fewer loops,
and because a condensed boson has less energy than an uncondensed boson,
the heat capacity decreases.




\end{enumerate}


\end{document}